\documentclass[11pt, epsf]{amsart}
\usepackage{epsfig}
\usepackage{eucal}
\usepackage{subfig}
\usepackage{amssymb}
\usepackage{latexsym}
\usepackage{amsfonts}
\usepackage[latin1]{inputenc}
\usepackage{slashbox}
\usepackage{colortbl}
\usepackage{mathrsfs}

\newenvironment{namelist}[1]{%
\begin{list}{}
 {
   
   \settowidth{\labelwidth}{#1}
   \setlength{\leftmargin}{1.1\labelwidth}
  }
 }{%
\end{list}}
\newcommand{\bi}{\begin{namelist}}
\newcommand{\ei}{\end{namelist}}

\newtheorem{Theo}{Theorem}

\newtheorem{Lem}{Lemma}
\newtheorem{Rem}{Remark}

\title{Stabilized Finite Element Method For The Radial Dirac Equation}
\author{Hasan Almanasreh, Sten Salomonson, and Nils Svanstedt}
\thanks{Department of Mathematical Sciences and Department of Physics,  University of Gothenburg,
        SE-412 96 Göteborg, Sweden}
\date{\today}

\begin{document}

\maketitle

\begin{abstract} A challenging difficulty in solving the radial Dirac eigenvalue problem numerically is the presence of spurious (unphysical) eigenvalues among the correct ones that are neither related to mathematical interpretations nor to physical explanations. Many attempts have been made and several numerical methods have been applied to solve the problem using finite element method (FEM), finite difference method (FDM), or other numerical schemes. Unfortunately most of these attempts failed to overcome the difficulty. As a FEM approach, this work can be regarded as a first promising scheme to solve the spuriousity problem completely. Our approach is based on an appropriate choice of trial and test functional spaces. We develop a Streamline Upwind Petrov-Galerkin method (SUPG) to the equation and derive an explicit stability parameter.\\\\

\end{abstract}

\textbf{Introduction.}
\\

Studying the properties of electrons in atoms is governed by the Dirac equation which gives a complete picture of the electron behavior by means of specifying its energies (eigenvalues) in orbitals around the nucleus. Up to date, computing the eigenvalues of an electron in the many-electron system (in some methods) is based on determining the corresponding eigenvalues in a single-electron system (Hydrogen-like ions), where the eigencouples are used as a basis to approximate the electron energies in the entire system. Unfortunately, computing the eigenvalues of the electron in the Hydrogen-like ions by numerical methods is upset by the presence of spurious solutions (eigenvalues do not match what is physically observed). The spurious solutions annoy the computations, they disturb the solution in a way it becomes no longer reliable. At the time, one can identify the spurious eigenvalues, but there is no efficient method to just remove them from the entire spectrum without affecting the genuine values.

The presence of spurious values in the spectrum of the radial Dirac equation and other problems has been addressed in most of numerical computations. In  \cite{PES}, the occurrence of the spurious roots has been related to incorrect balancing of large and small components spaces $\mathbf{H}^f$ and $\mathbf{H}^g$, and has been restricted to the positive quantum number $\kappa$.  In solving Dirac equation by a mapped Fourier grid \cite{ACK}, spurious values have been detected for $\kappa\geq 1$, where the causality is recounted to the symmetric treatment of the large and small components. For eigenvalue problems in general \cite{ZHA}, the occurrence of spectrum pollution has been related to the absence of suitable constraints in the mathematical formulations or discretization, which results in mismatching of desired physical properties of the problem. Shabaev and Tupitsyn et al. \cite{SHAB, TUPS} have also allied the presence of spectrum pollution to the symmetric discretization of the small and large components of the wave function.  They have pointed out that using the same finite space for both components is the essence of the problem. They have proposed an alternative method to handle the difficulty by an addition of suitable terms to the basis functions known as a basis correction. Also they have explained the property of energy coincidence for the positive and the corresponding negative values of $\kappa$.

The spuriousity of the eigenvalues computation using spectral Tau method has been studied in \cite{DAW}. Also the causality of the spurious solution in the electromagnetic problems in general can be found in \cite{MUR}. To the Dirac eigenvalue problem, we refer respectively to \cite{SAL} and \cite{FIS, SHAB} for finite difference and B-splines approximations. For a brief Finite Element derivation for the Dirac operator see e.g \cite{MUL}.\\

In the present work, we provide a finite element scheme for solving the radial Coulomb-Dirac operator that provides a complete treatment of the spurious eigenvalues. This scheme may be considered as the first stable finite element approach for the numerical approximation of the Dirac eigenvalue problem. To proceed, we relate the occurrence of spuriousity to the function spaces in the implemented numerical method. What ever the method is, finite element method (FEM), finite difference method (FDM), the spectral domain approach (SDA), the boundary element method (BEM), or the point matching method (PMM), the spuriousity persists. Hence, it is priorly understood as not an effect of the numerical method applied, but to a mismatching of some physical properties of the eigenstates in the computations. The present work interprets the existence of spurious values and their remedy by means of the following two steps:
\vspace{-0.01cm}
\begin{itemize}
\item [(1)] The choice of suitable trial functional space that meets the physical property of the wave functions in the implemented numerical methods and its role of spuriousity elimination.
\item [(2)] The choice of weighted test functional space, this treats what remains of spurious values in one hand, and solves the coincidence of energies for positive and corresponding negative $\kappa$ in the other.
\end{itemize}
\vspace{-0.01cm}
In other words, we classify the spurious solution of Dirac eigenvalue problem into two categories. The first is those that appear within the spectrum for all values of $\kappa$. We call this type the instilled spurious values. It is worth to mention that this type of spuriousity appears not only for positive $\kappa$, but for negative $\kappa$ as well. Instilled spurious values affect the true values or may degenerate with them which results in some perturbed eigenfunctions. However, this will be discussed in detail in the coming section, where, by means of choosing an appropriate space of discretization, part of the instilled spurious values is treated. The second category can be understood as the coincidence of the first eigenvalue of the radial operator with positive $\kappa$ to that with the corresponding negative $\kappa$. We call this type of spuriousity the unphysical coincidence phenomenon: The eigenvalues with positive $\kappa$ have been shown in the finite dimensional spaces to be a repetition for those with the corresponding negative $\kappa$ \cite{TUPS}, which is not the case in the usual (infinite) space of the wave functions. In an attempt to overcome the difficulty, the last (main) section is devoted to set a scheme that removes the spuriousity for both categories.

For a brief sketch of the scheme, consider the radial Coulomb-Dirac equation 
\begin{equation*}
\underbrace{\left(
\begin{array}{cc}
mc^2+V(x) & c\big(\!-\!D_x+\frac{\kappa}{x}\big) \\
c\big(D_x+\frac{\kappa}{x}\big) & -mc^2+V(x)
\end{array}
\right)}_{H_{\!r}}
\underbrace{ \left(
\begin{array}{c}
f(x) \\
g(x)
\end{array}
\right)}_{\varphi}
=
 \lambda \underbrace{\left(
\begin{array}{c}
f(x) \\
g(x)
\end{array}
\right)}_{\varphi}\, .
\end{equation*}

Here $m$ and $c$ are respectively the mass of the electron and the speed of light, the quantum number $\kappa$ is the spin-orbit coupling parameter defined as $\kappa\!=\!(-1)^{j+l+\frac{1}{2}}(j+\frac{1}{2})$, where $j$ and $l$ are the total and the orbital angular momentum numbers respectively, and $D_x$ is the derivative with respect to $x$ in $\mathbb{R}$. The Coulomb potential, $V(x)$, is given by $\frac{-Z}{x}$, where $Z\in[1\,,\,137]$ is the electric charge number. The unknown $f$ and $g$ are the large and small components of the eigenfunction $\varphi$ with corresponding eigenvalue $\lambda$.

The presence of convection terms in the off diagonal and the absence of diffusion terms cause numerical instability while computing the eigenvalues $\lambda$. Indeed, in the standard Galerkin finite element solution of the equation one encounters spurious eigenvalues. In order to remove the spuriousity, we derive a stable finite element scheme based on appropriate choice of functional spaces of the Dirac spinors. By rewriting the explicit equations of $f$ and $g$ and applying suitable boundary conditions, the original space of the Dirac wave functions is $\mathcal{H}_0(\Omega)=\{v\in C^1(\Omega)\cap H^1_0(\Omega): v'|_{\partial\Omega}\!=\!0\}$, where $C^1$ is the space of continuous functions which possess continuous first derivatives, $\Omega$ is some open bounded domain, and $H^1_0(\Omega)=\{v: v \text{ and }v'\text{ are elements of } L^2(\Omega),\text{ and }v|_{\partial\Omega}=0\}$  (for all values of $\kappa$ except $\pm1$, where for $\kappa=\pm1$ the lower boundary condition of $v'$ should differ from zero, but for generality and for sake of simplicity it is assumed to vanish, see Remark 1 below). Thus, by this definition, $\mathcal{H}_0(\Omega)$ is the space of continuous functions, $v$, which admit continuous first derivatives that are vanishing smoothly on the boundaries.

Consider the weak form of the equation above of finding $\{\lambda,\varphi\}\in \mathbb{R}\times \mathcal{H}_0(\Omega)^2 $ such that
\begin{equation*}
\int_{\Omega}\mathfrak{v}^t H_{\!r} \varphi dx=\lambda\int_{\Omega}\mathfrak{v}^t \varphi dx\,.
\end{equation*}
Where $\mathfrak{v}$ is a test function, and the superscript $t$ is the usual matrix transpose. Cubic Hermitian (CH) interpolation functions turn out to be a suitable choice which sufficiently fulfill the requirements of $\mathcal{H}_0(\Omega)$. Let $\mathcal{V}^H_h$ be the finite dimensional subspace of $\mathcal{H}_0(\Omega)$ on the partition $k_h$ spanned by the piecewise CH basis functions. Choosing $\mathfrak{v}\in (\mathcal{V}_h^H)^2$ as $(v,0)^t$ and $(0,v)^t$, where $v$ is an element of $\mathcal{V}_h^H$, and assuming  $f\;,g\in \mathcal{V}_h^H$, remove partially the first category of spuriousity (only for very small $Z$) and do not help in solving the coincidence phenomenon.

A complete treatment is achieved by letting the test function to live in another space different from that of the trial function, mainly by assuming $\mathfrak{v}$ to be $(v,\tau v')^t$ and $(\tau v', v)^t$ (where $v'$ means $D_xv$) respectively in the variational form above. The scheme is accomplished by deriving the stability parameter $\tau$, which turns out to have the form $ \tau:=\tau_j\cong \frac{9}{35}h_{j+1}\big(\frac{h_{j+1}-h_{j}}{h_{j+1}+h_{j}}\big)$. The derivation is based on two leading simplifications; to consider the limit operator in the vicinity of $x$ at infinity (i.e to consider the most numerically unstable part of the operator) and $c$-correspondence dominant parts of the system. From the weak form with the modified test function, and after applying the above simplifications we obtain an approximation $\lambda(\tau)$ of the accumulation eigenvalue. Knowing that the limit point eigenvalue is $mc^2$, we like to minimize the error $|\lambda(\tau)-mc^2|$, which gives the desired formula of $\tau_j$.\\

As a numerical method implemented in this work, the finite element method (FEM) is applied, with the usual continuous Galerkin method in the first section and a Petrov-Galerkin method in the second section. For the integrals evaluation, four-point Gaussian quadrature rule is applied.\\

The paper is arranged as follows; In the first part we discuss the first category of the spurious values and how to remove them partially via choosing suitable trial functional space. A comparison is performed between the incorrect and the correct functional spaces through numerical examples. In the second, we discuss the completion of the treatment. Basically we impose the weighted test function to live in a space different from that of the trial function. This is the well-known Streamline Upwind Petrov-Galerkin (SUPG) method \cite{ALM, DES, IDE}. Finally a stability parameter is derived to achieve the desired goal.
\section{Trial functional space}
Recall the radial Dirac equation
\begin{equation}\label{14}
\left(
\begin{array}{cc}
mc^2+V(x) & c\big(\!-\!D_x+\frac{\kappa}{x}\big) \\
c\big(D_x+\frac{\kappa}{x}\big) & -mc^2+V(x)
\end{array}
\right)
 \left(
\begin{array}{c}
f(x) \\
g(x)
\end{array}
\right)
=
 \lambda \left(
\begin{array}{c}
f(x) \\
g(x)
\end{array}
\right)\, ,
\end{equation}
where, then, the two-equation system is
\begin{equation}\label{15}
\big(mc^2+V(x)\big)f(x)+c\big(-g'(x)+\frac{\kappa}{x}g(x)\big)=\lambda f(x)\, ,
\end{equation}
\begin{equation}\label{16}
c\big(f'(x)+\frac{\kappa}{x}f(x)\big)+\big(-mc^2+V(x)\big)g(x)=\lambda g(x)\, .
\end{equation}

We first solve this system by usual continuous linear basis functions (hat functions). Since $x$ ranges over $[0 \,,\, \infty)$,  $x=0$ represents a singularity for the Coulomb potential and hence careful treatment is needed, i.e one can consider extended nucleus in the entire domain or just assume point nucleus on a cut-off domain. However, computationally, almost the same technique is used for both cases. For simplicity we will treat point nucleus model in all computations except in the last table where we extend the computations to extended nucleus.

Divide the domain $\Omega=[a \,,\, b]$ into $n+1$ subintervals with $n$ interior points distributed exponentially, and assume $k_h : a=x_0<x_1<x_2<\cdots <x_{n+1}=b$ the resultant partition of $\Omega$ with mesh size $h_i=x_i-x_{i-1}$.

The exponential distribution of the nodal points is crucial for solving the radial Dirac equation in order to get more nodal points near the singularity ($x=0$). This is because the wave function oscillates much more near the nucleus which means more information is needed about its behavior near that region, whereas the fine grid is not required in a position away from the nucleus.

The choice of the computational space $\mathcal{V}$ is important and plays the most influential role in the core of the problem. To see that, let us first take the space of only continuous functions as the functional space $\mathcal{V}$. We will show, by means of numerical examples, how this space results in the occurrence of spurious values. The presence of spuriousity is due to the fact that the only continuous functional space lacks to a certain constraint in the mathematical formulation, i.e it fails to have an identified property which being exist for the original wave function.

For a fast and simple algorithm, continuous linear basis functions are considered. So let $\mathcal{V}=\mathcal{V}^l$ be the subspace of continuous linear polynomials (the superscript $l$ denotes for the linear case), and let $\mathcal{V}_h^l\subset \mathcal{V}^l$ be the finite subspace consists of piecewise continuous linear polynomials on $k_h$ spanned by the usual linear functions. We assume that both trial and weighted test functions belong to this space. For $f(x)$ and $g(x)$ in $\mathcal{V}_h^l$ we write
\begin{eqnarray}\label{19}
f(x)=\sum_{j=1}^{n}\zeta_j\phi_j(x)\, ,\\
g(x)=\sum_{j=1}^{n}\xi_j\phi_j(x)\, ,
\end{eqnarray}
where $\zeta_j$ and $\xi_j$ are the unknown values of the functions $f$ and $g$ at the nodal point $x_j$ respectively, and $\phi_j$ is the basis function. Since the eigenfunction decays in the vicinity of $x$ at infinity and also considered to be zero at $x=0$, the Dirichlet conditions are assumed to treat the boundaries. The problem is now read as solving (\ref{15}) and (\ref{16}) such that $f=0$ and $g=0$ at $x=a\,,\,b$ (i.e $\zeta_0=\zeta_{n+1}=\xi_0=\xi_{n+1}=0$). The usual finite element method of the problem is to assume $f$ and $g$ as above in (\ref{15}) and (\ref{16}), then multiply by a test function and integrate over the domain $\Omega$
\begin{equation}\label{21}
\sum_{j=1}^n\big(w^{+}(x)\phi_j(x)\, ,\, v(x)\big)\zeta_j\,+\,\sum_{j=1}^n\big(-c\phi_j'(x)+\frac{\kappa}{x}\phi_j(x)\, , \, v(x)\big)\xi_j=\lambda\sum_{j=1}^n\big(\phi_j(x)\, ,\, v(x)\big)\zeta_j
\end{equation}
and
\begin{equation}\label{22}
\sum_{j=1}^n\big(c\phi_j'(x)+\frac{\kappa}{x}\phi_j(x)\, , \, v(x)\big)\zeta_j \,+\,\sum_{j=1}^n\big(w^{-}(x)\phi_j(x)\, ,\, v(x)\big)\xi_j=\lambda\sum_{j=1}^n\big(\phi_j(x)\, ,\, v(x)\big)\xi_j\, ,
\end{equation}
where $w^{\pm}(x)=\pm mc^2+V(x)$, and $\big(u\,,\,v\big)=\int_{\Omega}u\,v\,dx$ is the usual $L^2(\Omega)$ inner product. The basis function $\phi_j(x)$ has its support in $[x_{j-1}\,,\, x_j]=:I_j$ and  $[x_{j}\,,\, x_{j+1}]=I_{j+1}$ and defined as
\begin{displaymath}
\phi_j(x) = \left\{ \begin{array}{ll}
\frac{x-x_{j-1}}{h_j}& x\in I_j\, , \\
\frac{x_{j+1}-x}{h_{j+1}}& x\in I_{j+1}\, .
\end{array} \right.
\end{displaymath}

Let $v=\phi_i$ be an element of the same space $\mathcal{V}^l_h$ in (\ref{21}) and (\ref{22}), this leads to the symmetric generalized eigenvalue problem
\begin{equation}\label{23}
AX=\lambda BX\, .
\end{equation}
Here $A$ and $B$ are both symmetric block matrices defined by
\begin{equation}\label{24}
A=\left(\begin{array}{c|c}
mc^2M_{000}+M_{000}^V & -cM_{010}+c\kappa M_{001} \\
\hline
cM_{010}+c\kappa M_{001} & -mc^2M_{000}+M_{000}^V
\end{array}\right)
\end{equation}
and
\begin{equation}\label{25}
B=\left(\begin{array}{c|c}
M_{000}& 0 \\
\hline
0 & M_{000}
\end{array}\right)\, ,
\end{equation}
where $M_{rst}^q$ are $n\times n$ matrices defined as
\begin{equation}\label{26}
(M_{rst}^q)_{ij}=\int_{\Omega}\phi_j^{(s)}\,\phi_i^{(r)}\,x^{-t}\,q(x)\,dx\; ,\;\;\;\;\Big(\phi^{(r)}(x)=\frac{d^r}{dx^r}\phi (x)\Big)\, .
\end{equation}
The vector $X$ is the unknown defined as $(\zeta\, ,\, \xi)^t$, where  $$\zeta=(\zeta_1,\zeta_2,\cdots,\zeta_n)$$ and
$$\xi=(\xi_1,\xi_2,\cdots,\xi_n)\, .$$
Clearly the diagonal matrices of $A$ are symmetric and the off diagonal matrices consist of two parts, one is symmetric and exists in both sides, and the other, $(M_{010})^t=-M_{010}$, is anti-symmetric and exists in both off diagonal sides with different sign, this explains the symmetry of the block matrix $A$. For the block matrix $B$ the symmetry is obvious.

In Table 1 the first six computed eigenvalues for the Hydrogen atom $(Z=1)$ are listed for $|\kappa|=1$, these eigenvalues are obtained using $n=100$ interior nodal points. The exact solution for $\kappa=-1$ is shown in the right column of the table. Even with mesh refinement the spuriousity is still present, see Table 2 with $n=400$.
\begin{table}[h]
\begin{normalsize}
\caption{The first six computed eigenvalues for the electron in the Hydrogen atom using linear basis functions with $100$ nodal points.}
\centering
\begin{tabular}{@{} l c c r @{}}
\hline\hline
Level & $\kappa=1$ & $\kappa=-1$ & Rel. Form. $\kappa=-1$ \\ [0.1ex]
\hline
1 &\cellcolor[gray]{0.6} -0.50000665661 & -0.50000665659 & -0.50000665659 \\
2 & -0.12500414297 & -0.12500414298 & -0.12500208018 \\
3 & -0.05556140476 & -0.05556140479 & -0.05555629517 \\
 $\Rrightarrow$    &\cellcolor[gray]{0.6} -0.03192157994 &\cellcolor[gray]{0.6} -0.03192157993 & Spurious Eigenvalue \\
4 & -0.03124489833 & -0.03124489832 & -0.03125033803 \\
5 & -0.01981075633 & -0.19810756319 & -0.02000018105 \\ [1ex]
\hline
\end{tabular}
\label{table:nonlin}
\end{normalsize}
\end{table}
\vspace{-0.4cm}
\begin{table}[h]
\begin{normalsize}
\caption{The first six computed eigenvalues for the electron in the Hydrogen atom using linear basis functions with $400$ nodal points.}
\centering
\begin{tabular}{@{} l c c r @{}}
\hline\hline
Level & $\kappa=1$ & $\kappa=-1$ & Rel. Form. $\kappa=-1$ \\ [0.5ex]
\hline
1 &\cellcolor[gray]{0.6} -0.50000665661 & -0.50000665659 & -0.50000665659 \\
2 & -0.12500208841 & -0.12500208839 & -0.12500208018 \\
3 & -0.05555631532 & -0.05555631532 & -0.05555629517 \\
 $\Rrightarrow$    &\cellcolor[gray]{0.6} -0.03141172061 &\cellcolor[gray]{0.6} -0.03141172060 & Spurious Eigenvalue \\
4 & -0.03118772526 & -0.03118772524 & -0.03125033803 \\
5 & -0.01974434510 & -0.01974434508 & -0.02000018105 \\ [1ex]
\hline
\end{tabular}
\label{table:nonlin}
\end{normalsize}
\end{table}

In the tables above, the shaded left corner value is what meant by the unphysical coincidence phenomenon, and the values in the fourth row are the so-called instilled spuriousity. The spurious values appear for both positive and negative values of quantum number $\kappa$, and they persist despite of mesh refinement. Generally this kind of spurious solution can be identified among the right spectrum, but there is no way to just exclude them as a hope of treatment, since they have already affected or degenerated the true values.

As we mentioned before, the occurrence of spuriousity is related to the implementation of the numerical method, where the numerical scheme we assumed is the FEM with the proposed space $\mathcal{V}^l$. Therefore, either of them holds the responsibility of causing the spectrum pollution. At this end, it is worthy to mention that other methods like finite difference method (FDM), the method of moments (MoM) \cite{PES, SCH} and others, reported the occurrence of spuriousity in many computations for the Dirac operator or else. So we conclude that the problem of spuriousity is almost caused by the finite element spaces employed in the discretization, and hence the causality of spuriousity is $\mathcal{V}^l$-problem and never FEM-problem.

We return to (\ref{15}) and (\ref{16}), rewrite both equations to obtain explicit formulae for $f$ and $g$
\begin{eqnarray}\label{27}
w^{+}(x)\big(w^{-}(x)-\lambda\big)^2f(x)-\frac{c\kappa}{x}\big(w^{-}(x)-\lambda\big)\big(c f'(x)
+\frac{c\kappa}{x}f(x)\big)+c\Big[\big(w^{-}(x)-\lambda\big)\times \\ \big(c f''(x)+\frac{c\kappa}{x}f'(x)
-\frac{c\kappa}{x^2}f(x)\big)-V'(x)\big(c f'(x)+\frac{c\kappa}{x}f(x)\big)\Big]=\lambda\big(w^{-}(x)-\lambda\big)^2f(x)
\nonumber
\end{eqnarray}
\vspace{-0.08cm}
and
\vspace{-0.08cm}
\begin{eqnarray}\label{28}
w^{-}(x)\big(w^{+}(x)-\lambda\big)^2g(x)+\frac{c\kappa}{x}\big(w^{+}(x)-\lambda\big)\big(c g'(x)
-\frac{c\kappa}{x}g(x)\big)+c\Big[\big(w^{+}(x)-\lambda\big)\times \\ \big(c g''(x)-\frac{c\kappa}{x}g'(x)
+\frac{c\kappa}{x^2}g(x)\big)-V'(x)\big(c g'(x)-\frac{c\kappa}{x}g(x)\big)\Big]=\lambda\big(w^{+}(x)-\lambda\big)^2g(x)\, .
\nonumber
\end{eqnarray}
Equations (\ref{27}) and (\ref{28}) can be written in simpler forms as
\begin{equation}\label{29}
f''(x)+\gamma_1(x,\lambda)f'(x)+\gamma_2(x,\lambda)f(x)=0\, ,
\end{equation}
\begin{equation}\label{30}
g''(x)+\theta_1(x,\lambda)g'(x)+\theta_2(x,\lambda)g(x)=0\, .
\end{equation}
Where
$$
\gamma_1(x,\lambda)=-\frac{V'(x)}{w^{-}(x)-\lambda}\, ,
$$
$$
\gamma_2(x,\lambda)=\frac{\big(w^{+}(x)-\lambda\big)\big(w^{-}(x)-\lambda\big)}{c^2}-\frac{\kappa^2+
\kappa}{x^2}-\frac{\kappa V'(x)}{x\big(w^{-}(x)-\lambda\big)}\, ,
$$
$$
\theta_1(x,\lambda)=-\frac{V'(x)}{w^{+}(x)-\lambda}\, ,
$$
and
$$
\theta_2(x,\lambda)=\frac{\big(w^{+}(x)-\lambda\big)\big(w^{-}(x)-\lambda\big)}{c^2}-\frac{\kappa^2-
\kappa}{x^2}+\frac{\kappa V'(x)}{x\big(w^{+}(x)-\lambda\big)}\, .
$$

The terms $f''$ and $g''$ in (\ref{29}) and (\ref{30}) propose further constraint on both components of the wave function. By these equations $f$ and $g$ are imposed to be twice differentiable. This means that $f$ and $g$ should be continuous with continuous first derivatives, hence the proposed original domain is $C^1(\Omega)\cap H^1_0(\Omega)$.

Instead of regarding $\mathcal{V}^l$ as the space of variation, a space of continuous functions with continuous first derivative is considered to discretize both components of the wave function. At this end, one can think about a suitable space which meets the properties of $f$ and $g$; Lagrange interpolation functions are not suitable in this situation, since their first derivatives do not match the continuity property. So we consider instead a type of Hermitian functions (known as a generalization of the Lagrange functions) which are continuous and admit continuous first derivative.

The boundary conditions need special concern, they play a crucial role of choosing the space of discretization; Since the wave functions are assumed to vanish at the boundaries and by the smooth property of these functions, the way they move toward the boundaries should be in damping manner, i.e with vanishing velocity, this implies zero derivative boundary conditions should be considered as well (except the case when $\kappa=\pm1$ at the lower boundary, see Remark 1 below). Physically this is clearly reasonable, since the electron is neither expected to be close to the nucleus nor escaping to infinity.

Cubic Hermite (CH) interpolation functions turn out to be sufficient to fulfill the requirements. Such functions are third-degree piecewise polynomials consisting of two control points and two control tangent points for the interpolation. That means there is a control for both the function values and the derivatives at each nodal point $x_i$.

To study CH functions, let us first introduce the following spaces
\begin{itemize}
\item $\mathcal{H}(\Omega)=C^1(\Omega)\cap H^1_0(\Omega)$\,.
\item $\mathcal{H}_0(\Omega)=\{v\in \mathcal{H}(\Omega):  v'|_{\partial\Omega}\!=\!0\}$\,.
\end{itemize}
\begin{Rem}
\emph{
\begin{itemize}
\item [$(i)$] For the states $1s_{1/2}$ and $2p_{1/2}$ ($\kappa\!=\!-1$ and 1 respectively), the boundary conditions for the derivative of the components of the wave function are partially different, specifically at the lower boundary. I.e if $\partial \Omega^{up}$ and $\partial \Omega^{lo}$ denote respectively the upper and the lower boundaries, then $v'|_{\partial \Omega^{up}}\!=\!0\text{ and }v'|_{\partial \Omega^{lo}}\!\neq\!0$. This is due to the fact that the corresponding wave function do not vanish in a damping way near the origin, see \cite{SAL} for more details. Thus, for $\kappa\!=\!\pm1$, the same functional space $\mathcal{H}_0(\Omega)$ is considered but with small modification on the functions derivative at the lower boundary. Here we will keep the same notation $\mathcal{H}_0(\Omega)$ for the space for all $\kappa$'s, but when we mean the cases $\kappa\!=\!\pm1$ the right derivative condition at $\partial \Omega^{lo}$ should be considered.
\item [$(ii)$] For the sake of simplicity and as a matter of comparison, in the following computations of the energies of the electron in the Hydrogen atom, we do not use the right lower boundary conditions for $\kappa\!=\!\pm1$ as stated above. Instead we just assume zero for the derivative of the components of the wave function at $\partial \Omega^{lo}$, where the result might be slightly changed but does not affect the essence of the comparison. Also, without loss of generality, from now on we will assume $v'|_{\partial \Omega}\!=\!0$ for all $\kappa$.
\end{itemize}
}
\end{Rem}
Let $\mathcal{V}^H_h$ be the finite dimensional subspace of $\mathcal{H}_0$ on the partition $k_h$ spanned by CH basis functions. To summarize, $\mathcal{V}^H_h$ possesses the following properties:
\begin{itemize}
\item [$(i)$] It is a set of continuous piecewise CH polynomials.
\item [$(ii)$] $\forall v\in \mathcal{V}^H_h$, $v'$ exists and piecewise continuous.
\item [$(iii)$] $\forall v\in \mathcal{V}^H_h$, $v|_{\partial supp(v)}=v'|_{\partial supp(v)}=0$, where $\partial supp(v)$ is the boundaries of support of $v$.
\item [$(iv)$] It is a finite dimensional vector space of dimension $2n$ with basis $\{\phi_{j,1}\}_{j=1}^n$ and $\{\phi_{j,2}\}_{j=1}^n$ given below.
\end{itemize}

To approximate a function $u_h\in \mathcal{V}^H_h$, where the same partition $k_h$ of the same distribution is considered as before, $u_h$ can be written as
\begin{equation}\label{31}
u_h=\sum_{j=1}^n\xi_j\phi_{j,1}+\sum_{j=1}^n\xi_j'\phi_{j,2}\, ,
\end{equation}
$\xi_j$ and $\xi_j'$ are the unknown value of the function and its corresponding derivative at the nodal points $x_j$ respectively, and $\phi_{j,1}$ and $\phi_{j,2}$ are the basis functions of the space $\mathcal{V}_h^H$ having the following properties
\begin{displaymath}
\phi_{j,1}(x_i) = \left\{ \begin{array}{ll}
1\; ,& \text{If}\;j=i\, , \\
0\; ,& \text{Otherwise}\, ,
\end{array} \right.
\end{displaymath}
\begin{displaymath}
\phi_{j,2}'(x_i) = \left\{ \begin{array}{ll}
1\; ,& \text{If}\;j=i\, , \\
0\; ,& \text{Otherwise}\, ,
\end{array} \right.
\end{displaymath}
and
\begin{displaymath}
\phi_{j,1}'(x_i)=\phi_{j,2}(x_i)=0\;\; \forall i=1,2,\cdots n.
\end{displaymath}
It follows from the conditions above that $\phi_{j,1}$ interpolates the function values whereas $\phi_{j,2}$ is responsible of the function derivatives at the nodal point $x_j$. For non-uniform mesh, $\phi_{j,1}$ and $\phi_{j,2}$ are given by the following formulae (see also Figure 1 below, where the two basis functions are depicted for uniform and nonuniform meshes)

\begin{eqnarray}\label{32,33}
  \phi_{j,1}(x) &=& \left\{ \begin{array}{ll}
\frac{1}{h_j^2}(x-x_{j-1})^2-\frac{2}{h_j^3}(x-x_{j-1})^2(x-x_j)\, ,& x\in I_j\, , \\
1-\frac{1}{h_{j+1}^2}(x-x_j)^2+\frac{2}{h_{j+1}^3}(x-x_j)^2(x-x_{j+1})\, ,& x\in I_{j+1}\, ,
\end{array} \right. \\
  \phi_{j,2}(x)&=&  \left\{ \begin{array}{ll}
\frac{1}{h_j^2}(x-x_{j-1})^2(x-x_j)\, ,& x\in I_j\, , \\
(x-x_j)-\frac{1}{h_{j+1}}(x-x_j)^2+\frac{1}{h_{j+1}^2}(x-x_j)^2(x-x_{j+1})\, ,& x\in I_{j+1}\, .
\end{array} \right.
\end{eqnarray}

\begin{figure}[h]
\centering
\includegraphics[width=6cm]{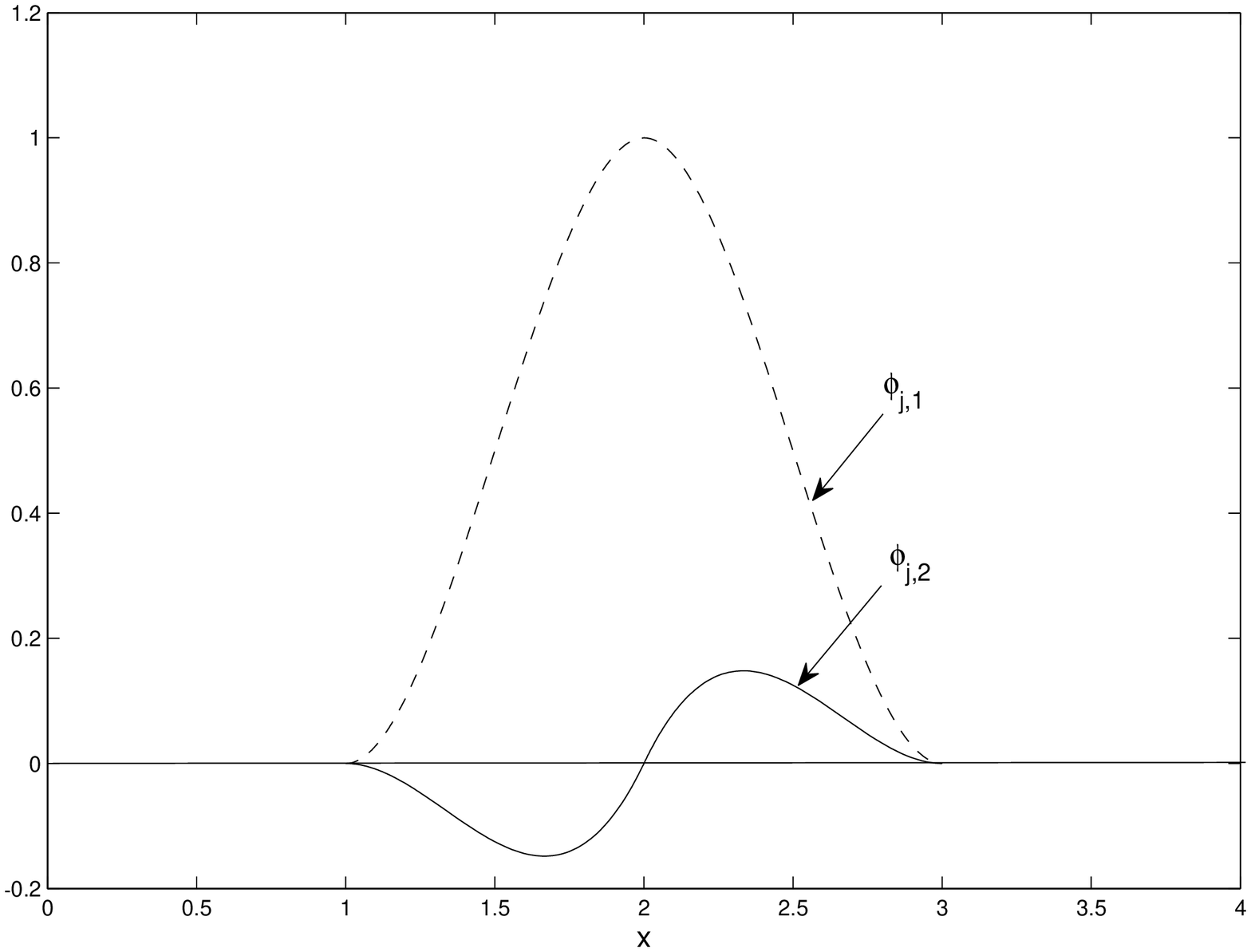}
\includegraphics[width=6cm]{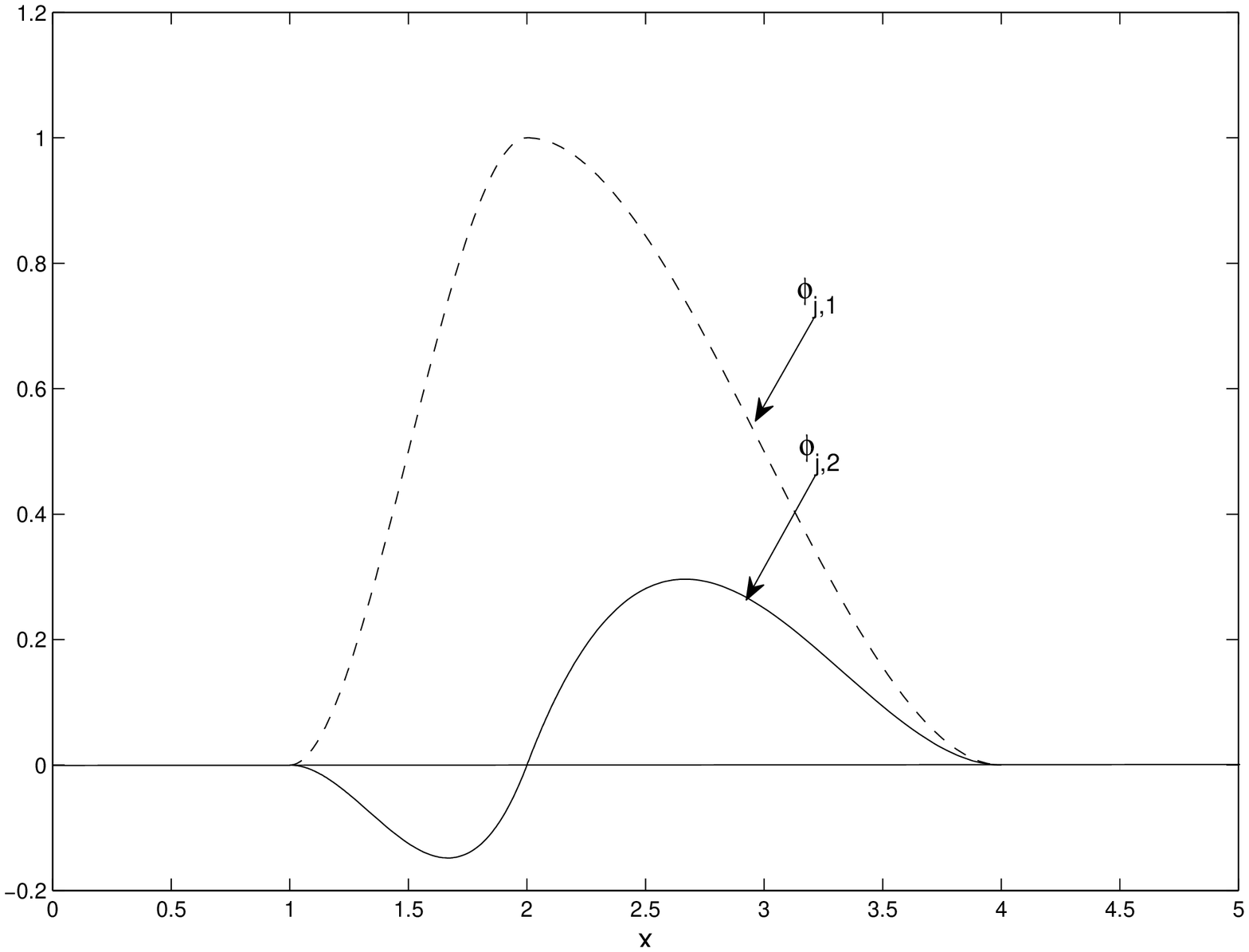}
\caption{CH basis functions with uniform distributed nodal points(Left) and nonuniform distributed nodal points(Right).}
\end{figure}
The approximation error using CH basis functions in the subinterval $I_j$ is given by
\begin{equation}\label{35}
|u-u_h|\leq c_1h^4\|u^{(4)}\|_{L^\infty (I_j)}\, ,
\end{equation}
where $c_1=\frac{1}{384}$, and $h=\max_{j}h_j$. So the error bound is obtained individually for each subinterval $I_j$, yielding a fine-grained error bound, which means that CH basis produces more accuracy compared to the linear or quadratic interpolation function in general.

To construct FEM for the radial Dirac equation using CH basis functions, we as usual multiply (\ref{15}) and (\ref{16}) by test function $v\in\mathcal{H}_0(\Omega)$ and integrate over $\Omega$. To discretize the system we assume $f$ and $g$ are elements of $\mathcal{V}_h^H$, thus they can be written as
\begin{eqnarray}\label{37}
f(x)=\sum_{j=1}^{n}\zeta_j\phi_{j,1}(x)+\sum_{j=1}^{n}\zeta_j'\phi_{j,2}(x)\, ,\\
g(x)=\sum_{j=1}^{n}\xi_j\phi_{j,1}(x)+\sum_{j=1}^{n}\xi_j'\phi_{j,2}(x)\, ,
\end{eqnarray}
where $\zeta_j$ and $\zeta_j'$ are the nodal value and the nodal derivative of $f$ respectively at $x_j$, and $\xi_j$ and $\xi_j'$ are the corresponding ones associated to $g$. This yields
\begin{eqnarray}\label{47}
\sum_{j=1}^n\Big(-c\phi_{j,1}'(x)+\frac{c\kappa}{x}\phi_{j,1}(x)\, ,\, v(x)\Big)\xi_j
+\sum_{j=1}^n\Big(-c\phi_{j,2}'(x)+\frac{c\kappa}{x}\phi_{j,2}(x)\, ,\, v(x)\Big)\xi_j'+\\
+\sum_{j=1}^n\Big(w^{+}(x)\phi_{j,1}(x)\, ,\, v(x)\Big)\zeta_j +\sum_{j=1}^n\Big(w^{+}(x)\phi_{j,2}(x)\, ,\, v(x)\Big)\zeta_j'\nonumber\\=
\lambda \Big[\sum_{j=1}^n\Big(\phi_{j,1}(x)\, ,\, v(x)\Big)\zeta_j +\sum_{j=1}^n\Big(\phi_{j,2}(x)\, ,\, v(x)\Big)\zeta_j'\Big]\, ,\nonumber
\end{eqnarray}
\begin{eqnarray}\label{48}
\sum_{j=1}^n\Big(c\phi_{j,1}'(x)+\frac{c\kappa}{x}\phi_{j,1}(x)\, ,\, v(x)\Big)\zeta_j
+\sum_{j=1}^n\Big(c\phi_{j,2}'(x)+\frac{c\kappa}{x}\phi_{j,2}(x)\, ,\, v(x)\Big)\zeta_j'+\\
+\sum_{j=1}^n\Big(w^{-}(x)\phi_{j,1}(x)\, ,\, v(x)\Big)\xi_j +\sum_{j=1}^n\Big(w^{-}(x)\phi_{j,2}(x)\, ,\, v(x)\Big)\xi_j'\nonumber \\=
\lambda \Big[ \sum_{j=1}^n\Big(\phi_{j,1}(x)\, ,\, v(x)\Big)\xi_j + \sum_{j=1}^n\Big(\phi_{j,2}(x)\, ,\, v(x)\Big)\xi_j' \Big]\, .\nonumber
\end{eqnarray}
Let $v(x)$ be an element of $\mathcal{V}_h^H$, and consider (\ref{47}) and (\ref{48}) first with $v(x)=\phi_{i,1}(x)$ and then with $v(x)=\phi_{i,2}(x)$. This yields the following system
\begin{equation}\label{49}
\mathcal{A}X=\lambda\mathcal{B}X\,,
\end{equation}
where
\begin{equation}\label{50}
\mathcal{A}=\left(\begin{array}{c|c}
mc^2M\negmedspace M_{000}+M\negmedspace M_{000}^V & -cM\negmedspace M_{010}+c\kappa M\negmedspace M_{001} \\
\hline
cM\negmedspace M_{010}+c\kappa M\negmedspace M_{001} & -mc^2M\negmedspace M_{000}+M\negmedspace M_{000}^V
\end{array}\right)
\end{equation}
and
\begin{equation}\label{51}
\mathcal{B}=\left(\begin{array}{c|c}
M\negmedspace M_{000}& 0 \\
\hline
0 & M\negmedspace M_{000}
\end{array}\right)\, .
\end{equation}
The vector $X$ is the unknown given by $X=(\zeta,\zeta',\xi,\xi')$, and the general block matrices $M\negmedspace M_{rst}^q$ are defined as
\begin{equation}\label{52}
M\negmedspace M_{rst}^q=\left(\begin{array}{c|c}
M_{rst(1,1)}^q& M_{rst(1,2)}^q \\
\hline
M_{rst(2,1)}^q & M_{rst(2,2)}^q
\end{array}\right)\, ,
\end{equation}
where
\begin{equation}\label{53}
(M_{rst(k,l)}^q)_{ij}=\int_{\Omega}\phi_{j,l}^{(s)}\,\phi_{i,k}^{(r)}\,x^{-t}\,q(x)\,dx\, . \end{equation}
Tables 3 and 4 contain the first six computed eigenvalues of the radial Dirac operator for the Hydrogen atom, with $n=100$ and $400$ interior nodal points using CH basis functions. The computation is run for $\kappa=-1,1$, and the right column represents the exact solution for $\kappa=-1$.
\vspace{-0.5cm}
\begin{table}[h]
\begin{normalsize}
\caption{The first six computed eigenvalues for the electron in the Hydrogen atom using CH basis functions with $100$ nodal points.}
\centering
\begin{tabular}{@{} l c c r @{}}
\hline\hline
Level & $\kappa=1$ & $\kappa=-1$ & Rel. Form. $\kappa=-1$ \\ [0.5ex]
\hline
1 &\cellcolor[gray]{0.6} -0.50000632471 & -0.50000665659 & -0.50000665659 \\
2 & -0.12500207951 & -0.12500207951 & -0.12500208018 \\
3 & -0.05555629341 & -0.05555629338 & -0.05555629517 \\
4 & -0.03125018386 & -0.03125018404 & -0.03125033803 \\
5 & -0.01982545837 & -0.01982545886 & -0.02000018105 \\
6 & -0.01085968925 & -0.01085968695 & -0.01388899674 \\ [1ex]
\hline
\end{tabular}
\label{table:nonlin}
\end{normalsize}
\end{table}
\begin{table}[h]
\begin{normalsize}
\caption{The first six computed eigenvalues for the electron in the Hydrogen atom using CH basis functions with $400$ nodal points.}
\centering
\begin{tabular}{@{} l c c r @{} }
\hline\hline
Level & $\kappa=1$ & $\kappa=-1$ & Rel. Form. $\kappa=-1$ \\ [0.5ex]
\hline
1 &\cellcolor[gray]{0.6} -0.50013790178 & -0.50000665659  & -0.50000665659 \\
2 & -0.12500208021 & -0.12500208018  & -0.12500208018 \\
3 & -0.05555629517 & -0.05555629518  & -0.05555629517 \\
4 & -0.03125027925 & -0.03125027916  & -0.03125033803 \\
5 & -0.01985891281 & -0.01985888664  & -0.02000018105 \\
6 & -0.01116648473 & -0.01116629119  & -0.01388899674 \\ [1ex]
\hline
\end{tabular}
\label{table:nonlin}
\end{normalsize}
\end{table}

It is noted, from the tables above, that some instilled spurious values are removed (the values that were present between level 3 and level 4). Also the speed of convergence to the exact eigenvalues is enhanced as the number of nodal points is increased. Unfortunately, part of the instilled spuriousity is still present for most values of $Z$, also the coincidence remains unsolved.

The unphysical coincidence phenomenon assigns almost the same energies for
\begin{eqnarray*}
 && N_{p^{\frac{1}{2}}}(\kappa=1)\;\;\text{and}\;\; N_{s^{\frac{1}{2}}}(\kappa=-1)\, , \;
N\geq 2. \\
  && N_{d^{\frac{3}{2}}}(\kappa=2)\;\;\text{and}\;\; N_{p^{\frac{3}{2}}}(\kappa=-2)\, , \;
N\geq 3. \\
   && N_{f^{\frac{5}{2}}}(\kappa=3)\;\;\text{and}\;\; N_{d^{\frac{5}{2}}}(\kappa=-3)\, , \;
N\geq 4.\quad\text{etc.}
\end{eqnarray*}

The occurrence of this phenomenon is deeply studied for both nonrelativistic and relativistic cases; In \cite{THA}, the coincidence of energies is proved for the same values of $\kappa$ that differ in sign via studying the commutation of Dirac operator with Biedenharn-Johnson-Lippmann (BJL) operator in the relativistic case. Also in the nonrelativistic case, the energy dependence on the quantum number $\varrho=n+|\kappa|$ is proved, which implies the energy independence of the sign of $\kappa$. The coincidence of the energies in the finite space is also studied in \cite{TUPS}, where the spuriousity in general is interpreted as an effect of the same treatment of both components of the wave functions.

As it is known that the exact solution of the Dirac operator with Coulomb potential for point nucleus results in different lowest bound energies for different values of $\kappa$. In this work, as it is pointed before, we relate the problem of energies coincidence to the numerical implementation. Roughly speaking not to the method of approximation, but to the proposed spaces of discretization.\\

In the last computations we imposed the test functions to live in the same space as well the trial functions, that is the usual Galerkin method. As we have seen, this results in a solution not cleaned from spurious values. However, it is well-known that the Galerkin method when it is applied to convection dominant problem, the solution will be upset by perturbations, which is getting worse with the increase in the convection size.

Nevertheless, it is assumed non uniform mesh (exponentially distributed nodal points) to match desirable requirements of high resolution near the nucleus, where the wave functions oscillate rapidly compared to their oscillations in a region away from it. This means that for each nodal point $x_j$ there are two adjacent systems of what are called fine-mesh grid with much larger coarse mesh. Hence when the wave function crosses the interface between these two regions, its phase is altered to fit the unbalanced change in the displacement size. One can understand the concept by regarding the variant mesh as different media to the generating waves, where most of those waves are not resolvable in two different meshes at the same time. We refer to \cite{CUL} and \cite{LON73, LON75} for more details.

Also, from numerical algebra point of view, one considers the linear system given by (\ref{49}) and posteriorly  notes that the sign of $\kappa$ that appears as a factor of the block matrix $M\negmedspace M_{001}$ does not contribute in determining the eigenvalues, which is entirely incorrect from physical point of view. So what is needed is to let the sign of $\kappa$ play a role in eigenvalues definition. This can be achieved by clever and justified addition of terms that includes $\kappa$ without deforming the original equations. These motivations suggest to use an alternative method to Galerkin formulation that does not demonstrate instability at the time treats the phenomenon of coincidence.

Streamline Upwind Petrov-Galerkin (SUPG) method is used to solve the problem, which consistently introduces additional stability terms in the upwind direction, these terms are based on the residual quantities for the governing equations and on the modification of the weighted test functional space. The latter is understood as adapting the test function $\mathfrak{v}$ from being $(v,0)$ and $(0,v)$ to be $(v,\tau v')$ and $(\tau v',v)$ respectively, so it is a type of residual corrections added to the original equations. Tau ,$\tau$, is called the stability parameter which we are investigating, where its derivation is the main part of the upcoming section.
\section{Weighted Functional Space}
To stabilize FEM approximation applied to the Dirac operator, modified SUPG is used to formulate the problem. This consists in adding suitable stability terms to the standard Galerkin method. The SUPG method is designed to maintain the consistency, so that the solution of the original problem is still a solution of the variational equations.

The idea behind SUPG is to introduce a diffusion term $(u'\, ,\,v')$ which eliminates the instability and enhances the approximation without modifying the problem. Several approaches can be implemented to create such term. To mention, we can just artificially add $(au'\, ,\,v')$, where $a$ is a constant that controls the diffusivity size, this method is first order accurate at most. Or the artificial diffusion term can be added in the direction of the streamlines to avoid excess diffusivity \cite{BRO81, BRO82}, even though this method introduces less crosswind diffusivity than the first mentioned, but it is still inconsistent modification. The methods mentioned above result in a modified problem differs from the original by the addition of the terms which alter the structure of the problem and force the exact solution to be no longer satisfying the variational equations.

To formulate the method, consider the radial Dirac equation
$$
H_{\!r}\varphi=\lambda\varphi\; ,\quad \text{where} \quad \varphi=\big(f(x)\, ,\, g(x)\big)^t\quad \text{and}
$$
$$
H_{\!r} = \left(
\begin{array}{cc}
w^{+}(x) & c\big(\!-\!D_x+\frac{\kappa}{x}\big) \\
c\big(D_x+\frac{\kappa}{x}\big) & w^{-}(x)
\end{array}
\right)\, ,
$$
which is equivalent to
\begin{equation}\label{56}
\left(
\begin{array}{c}
w^{+}(x)f(x)-cg'(x)+\frac{c\kappa}{x}g(x) \\
cf'(x)+\frac{c\kappa}{x}f(x)+w^{-}(x)g(x)
\end{array}
\right)
 =
 \lambda \left(
\begin{array}{c}
f(x) \\
g(x)
\end{array}
\right)\, .
\end{equation}
Define the residual functional of each equation as
\begin{equation}\label{57}
R^{1}_e\big(f,g\big)(x)=w^{+}(x)f(x)-cg'(x)+\frac{c\kappa}{x}g(x)-\lambda f(x)=\big(W^{+}f-cg'+\frac{c\kappa}{x}g\big)(x)\, ,
\end{equation}
\begin{equation}\label{58}
R^{2}_e \big(f,g\big)(x)=cf'(x)+\frac{c\kappa}{x}f(x)+w^{-}(x)g(x)-\lambda g(x)=\big(W^{-}g+cf'+\frac{c\kappa}{x}f\big)(x)\, .
\end{equation}
Here $W^{\pm}(x)=w^{\pm}(x)-\lambda$.\\
The previously derived Galerkin discretization with CH basis functions reads
\begin{equation}\label{59}
\int_{\Omega}\mathfrak{v}^t H_{\!r} \varphi dx=\lambda\int_{\Omega}\mathfrak{v}^t \varphi dx\, ,
\end{equation}
where $\mathfrak{v}$ is $(v,0)^t$ and $(0,v)^t$, and
\begin{equation}\label{60}
\varphi(x) = \left(
\begin{array}{c}
f(x) \\
g(x)
\end{array}
\right)
= \left(
\begin{array}{c}
\sum_{j=1}^{n}\zeta_j\phi_{j,1}(x)+\sum_{j=1}^{n}\zeta_j'\phi_{j,2}(x) \\
\sum_{j=1}^{n}\xi_j\phi_{j,1}(x)+\sum_{j=1}^{n}\xi_j'\phi_{j,2}(x)
\end{array}
\right)\, .
\end{equation}
So far with Galerkin approximation the components of $\mathfrak{v}$ as well as $f$ and $g$ are elements of $\mathcal{V}_h^H$.

SUPG is formulated based on modifying the test function $\mathfrak{v}$ to a form that includes $v'$ as a correction term to introduce the required diffusivity. Hence we assume $v\in \mathcal{V}_h^H$ as well $f$ and $g$, but  $\mathfrak{v}\notin (\mathcal{V}_h^H)^2$ is just continuous function. I.e, let $\mathfrak{v}$ be $(v,\tau v')^t$ and $(\tau v', v)^t$ in (\ref{59}), where $\tau$ is the stability parameter to be studied soon. This leads to
\begin{equation}\label{61}
\big(w^{+}f\, ,\, v\big)+\big(-cg'+\frac{c\kappa}{x}g\, ,\, v\big)+\big(R^2_e(f,g)\, ,\, \tau v'\big)=\lambda\big(f\, ,\, v\big)
\end{equation}
and
\begin{equation}\label{62}
\big(cf'+\frac{c\kappa}{x}f\, ,\, v\big)+\big(w^{-}g\, ,\, v\big)+\big(R^1_e(f,g)\, ,\, \tau v'\big)=\lambda\big(g\, ,\, v\big)\, ,
\end{equation}
Each of the discretizations above, using the new weighted test functions, is the usual Galerkin formulation with additional perturbation terms consist of the weak variational form of the residual of the opposite equation with basis function $\tau v'$. This keeps $f$ and $g$, the exact solution, satisfying the weak formulation without modifying the problem.

In matrix notations, the system $\mathcal{A}X=\lambda\mathcal{B}X$ is obtained as before, but $\mathcal{A}$ and $\mathcal{B}$ are slightly perturbed by additional matrices factored by $\tau$
\begin{equation}\label{63}
\mathcal{A}=\left(\begin{array}{c|c}
mc^2M\negmedspace M_{000}+M\negmedspace M_{000}^V+ & -cM\negmedspace M_{010}+c\kappa M\negmedspace M_{001}+\\
+c\tau M\negmedspace M_{110}+c\tau\kappa M\negmedspace M_{101} & -mc^2\tau M\negmedspace M_{100}+\tau M\negmedspace M_{100}^V\\
\hline
cM\negmedspace M_{010}+c\kappa M\negmedspace M_{001}+  & -mc^2M\negmedspace M_{000}+M\negmedspace M_{000}^V+\\
mc^2\tau M\negmedspace M_{100}+\tau M\negmedspace M_{100}^V & -c\tau M\negmedspace M_{110}+c\tau\kappa M\negmedspace M_{101}
\end{array}\right)
\end{equation}
and
\begin{equation}\label{64}
\mathcal{B}=\left(\begin{array}{c|c}
M\negmedspace M_{000} & \tau M\negmedspace M_{100} \\
\hline
\tau M\negmedspace M_{100} & M\negmedspace M_{000}
\end{array}\right)\, .
\end{equation}
The unknown vector $X$ and the generalized block matrices $M\negmedspace M_{rst}^q$ are as defined before. It is notable from the system above that the resultant block matrices $\mathcal{A}$ and $\mathcal{B}$ are not symmetric any more, in this situation complex eigenvalues may will begin to appear, which of course what we should avoid in the computations. To be more precise, the appearance of complex eigenvalues depends on the size of $\tau$, where they do appear for large size. For small size of $\tau$ one can consider the above system as the usual system that corresponds to the Galerkin approximation (which is symmetric) with an addition of small perturbation of size $\tau$, which still admits real eigenvalues.\\

Now, the main task is to determine the stability parameter $\tau$ that completes the scheme of removing the spuriousity for both categories and improves the convergence. The derivation $\tau$ assumes non full dependence on the exact solution of the complete operator for point nucleus, instead limit operator is assumed. Parallel with considering the dominant terms relative to the speed of light. Before proceeding into details, we will give some lemmas without complete proofs, where the proofs in some cases are simple.

The following lemma provides the approximated values of the radial function $f$ and $g$ at the nodal point $x_j$, where backward and forward derivative approximations are implemented, hence the error is $\mathcal{O}(h)$.
\begin{Lem}
For the Dirac spinors $f$ and $g$, let $\zeta_{j-1}$, $\zeta_{j+1}$, $\xi_{j-1}$, and $\xi_{j+1}$ be the $f$'s and $g$'s nodal values at $x_{j-1}$ and $x_{j+1}$ respectively, then the following holds
\begin{eqnarray*}
 && \zeta_{j-1}\cong \Big(1+\frac{h_j\kappa}{x_j}\Big)\zeta_j+\Big(\frac{h_j}{c}\big(-mc^2+V(x_j)\big)-\frac{h_j}{c}\lambda \Big)\xi_j\, . \\
  &&  \xi_{j-1}\cong \Big(1-\frac{h_j\kappa}{x_j}\Big)\xi_j+\Big(-\frac{h_j}{c}\big(mc^2+V(x_j)\big)+\frac{h_j}{c}\lambda \Big)\zeta_j\, . \\
   &&  \zeta_{j+1}\cong \Big(1-\frac{h_{j+1}\kappa}{x_j}\Big)\zeta_j+\Big(-\frac{h_{j+1}}{c}\big(-mc^2+V(x_j)\big)+
   \frac{h_{j+1}}{c}\lambda \Big)\xi_j\, . \\
    && \xi_{j+1}\cong \Big(1+\frac{h_{j+1}\kappa}{x_j}\Big)\xi_j+\Big(\frac{h_{j+1}}{c}\big(mc^2+V(x_j)\big)-
    \frac{h_{j+1}}{c}\lambda \Big)\zeta_j\, .
\end{eqnarray*}
\end{Lem}
\hspace{-4mm}\underline{\emph{Proof}}. Consider the two-equation system of the radial Dirac equation
\begin{eqnarray*}
\big(mc^2+V(x)\big)f(x)+c\big(-g'(x)+\frac{\kappa}{x}g(x)\big)=\lambda f(x)
\end{eqnarray*}
and
\begin{eqnarray*}
c\big(f'(x)+\frac{\kappa}{x}f(x)\big)+\big(-mc^2+V(x)\big)g(x)=\lambda g(x)\,.
\end{eqnarray*}
Assuming the above system for arbitrary $x_j\in k_h$, and using the backward and the forward difference approximations for the derivatives (backward $\Rightarrow f'|_{x_j}\cong \frac{f(x_j)-f(x_{j-1})}{x_j-x_{j-1}}=\frac{\zeta_j-\zeta_{j-1}}{h_j}$ and forward $\Rightarrow f'|_{x_j}\cong \frac{f(x_{j+1})-f(x_{j})}{x_{j+1}-x_{j}}=\frac{\zeta_{j+1}-\zeta_j}{h_{j+1}}$), one gets the desired results.\hfill{$\blacksquare$}\\

For the computed matrices $M\negmedspace M_{000}$, $M\negmedspace M_{100}$, $M\negmedspace M_{010}$, and  $M\negmedspace M_{110}$ in the block systems (\ref{63}) and (\ref{64}), the exact element integrals are obtained by the following lemma. For the remaining matrices one can calculate the exact element integrals, but it is rather hard to get them simplified. Therefore, we just point out in Remark 3 notations for the desired values without writing the explicit expressions.
\begin{Lem}
The following table is the exact element integrals for some matrices in the generalized system.
\vspace{-0.3cm}
\begin{tiny}
\begin{table}[h]
\caption{The exact element integrals for some matrices in the generalized system.}
\centering
\begin{tabular}{@{}|| l | c || c || c || c ||| c | c | c || @{}}
\hline\hline
\backslashbox{Matrix}{Index}& \backslashbox{Row}{Column}& $j-1$ & $j$ &$j+1$ & $j-1+n$ & $j+n$ & $j+1+n$ \\[1ex]\hline\hline
&$j$ &\cellcolor[gray]{0.6}$\frac{9}{70}h_{j+1}$ &\cellcolor[gray]{0.6} $\frac{13}{35}(h_{j+1}+h_j)$ &\cellcolor[gray]{0.6} $\frac{9}{70}h_{j+1}$ & $\frac{13}{420}h_{j+1}^2$ & $\frac{11}{210}(h_{j+1}^2-h_j^2)$ & $-\frac{13}{420}h_{j+1}^2$ \\[1ex]
\raisebox{1.4ex}{$M\negmedspace M_{000}$}&$j+n$
& $-\frac{13}{420}h_{j+1}^2$ & $\frac{11}{210}(h_{j+1}^2-h_j^2)$ & $\frac{13}{420}h_{j+1}^2$  & $-\frac{1}{140}h_{j+1}^3$  & $\frac{1}{105}(h_{j+1}^3+h_j^3)$ & $-\frac{1}{140}h_{j+1}^3$ \\[1ex]
\hline\hline
&$j$ &\cellcolor[gray]{0.6} $\frac{1}{2}$ & \cellcolor[gray]{0.6}$0$ & \cellcolor[gray]{0.6}$-\frac{1}{2}$ & $\frac{1}{10}h_{j+1}$ & $-\frac{1}{10}(h_{j+1}+h_j)$& $\frac{1}{10}h_{j+1}$ \\[1ex]
\raisebox{1.4ex}{$M\negmedspace M_{100}$}&$j+n$
&$-\frac{1}{10}h_{j+1}$ & $\frac{1}{10}(h_{j+1}+h_j)$ & $-\frac{1}{10}h_{j+1}$ & $-\frac{1}{60}h_{j+1}^2$ & $0$ &$\frac{1}{60}h_{j+1}^2$ \\[1ex]
\hline\hline
&$j$ & \cellcolor[gray]{0.6}$-\frac{1}{2}$ & \cellcolor[gray]{0.6}$0$& \cellcolor[gray]{0.6}$\frac{1}{2}$ & $-\frac{1}{10}h_{j+1}$ & $\frac{1}{10}(h_{j+1}+h_j)$ & $-\frac{1}{10}h_{j+1}$ \\[1ex]
\raisebox{1.4ex}{$M\negmedspace M_{010}$}&$j+n$
&$\frac{1}{10}h_{j+1}$& $-\frac{1}{10}(h_{j+1}+h_j)$ &$\frac{1}{10}h_{j+1}$ &$\frac{1}{60}h_{j+1}^2$ & $0$ & $-\frac{1}{60}h_{j+1}^2$ \\[1ex]
\hline\hline
&$j$ &\cellcolor[gray]{0.6}$-\frac{6}{5}\frac{1}{h_{j+1}}$& \cellcolor[gray]{0.6}$\frac{6}{5}\frac{h_{j+1}+h_j}{h_{j+1}h_j}$ &\cellcolor[gray]{0.6} $-\frac{6}{5}\frac{1}{h_{j+1}}$& $-\frac{1}{10}$ & $0$ & $\frac{1}{10}$ \\[1ex]
\raisebox{1.4ex}{$M\negmedspace M_{110}$}&$j+n$
&$\frac{1}{10}$ & $0$ & $-\frac{1}{10}$ & $-\frac{1}{30}h_{j+1}$ & $\frac{2}{15}(h_{j+1}+h_j)$ & $-\frac{1}{30}h_{j+1}$ \\[1ex]
\hline\hline
\end{tabular}
\end{table}
\end{tiny}
\end{Lem}
\vspace{-0.2cm}
\hspace{-4mm}\underline{\emph{Proof}}. The proof is straight forward by evaluating the integrals.\hfill{$\blacksquare$}
\begin{Rem}
\emph{
The basis functions consist of two parts, one corresponds to the function value and the other to the function derivative (the latter with no considerable contribution to the function values) at the nodal points. Therefore we will, for simplicity, just take into account the part of the basis functions that contributes to the function values at the nodal point only. Thus, the upper left (shaded) three-cell corner of each matrix of the above table is considered.
}
\end{Rem}
\begin{Rem}
\emph{
For the other matrices in the block matrix $\mathcal{A}$, $M\negmedspace M_{001}$ and $M\negmedspace M_{101}$ (where $M\negmedspace M_{000}^V$ and $M\negmedspace M_{100}^V$ can be written respectively as $-ZM\negmedspace M_{001}$ and $-ZM\negmedspace M_{101}$ for $V(x)=\frac{-Z}{x}$), we will use the following notations for the calculations of the element integral using the part of the basis functions that contributes only the function values at the nodal points as indicated in the remark above. Namely as a matter of notation we denote the following
\begin{table}[h]
\caption{The element integrals notations of the matrices $M\negmedspace M_{001}$ and $M\negmedspace M_{101}$ for the $j^{th}$ row.}
\centering
\begin{tabular}{@{} ||c|c|c|c|| @{} }
\hline\hline
\backslashbox{Matrix}{Index} & $j-1$ & $j$ & $j+1$ \\ [0.5ex]
\hline
$M\negmedspace M_{001}$ & $s_{j-1}$ & $s_j$ & $s_{j-1}$ \\
\hline
$M\negmedspace M_{101}$ & $r_{j-1}$ & $r_j$ &$ r_{j+1}$\\
 [1ex]
\hline
\hline
\end{tabular}
\end{table}
}
\end{Rem}

Now we are at the position to state the main theorem of the stability parameter $\tau$.
\begin{Theo}
The mesh-dependence stability parameter $\tau$ that appears in the formulations (\ref{61}) and (\ref{62}) is of the following form
\begin{equation}\label{65}
\tau:=\tau_j\cong \frac{9}{35}h_{j+1}\frac{(h_{j+1}-h_j)}{(h_{j+1}+h_j)}\, .
\end{equation}
\end{Theo}
Before proceeding, we introduce the following notations to ease handling the proof.
\begin{eqnarray*}
c_1 \!\!&\!=\!&\!\! -\frac{(h_{j+1}+h_j)}{2c}\;. \\
c_2 \!\!&\!=\!&\!\! -\frac{9}{70}(h_{j+1}-h_j)\;. \\
c_3 \!\!&\!=\!&\!\! \frac{9}{70}\frac{\kappa}{x_j}h_{j+1}(h_{j+1}-h_j)-\kappa s_{j-1}(h_{j+1}-h_j)-\frac{Z}{2cx_j}(h_{j+1}+h_j)\tau_j +\frac{Z}{c}(r_{j+1}h_{j+1}+\\
\!\!&\!\!&\!\!-r_{j-1}h_j)\tau_j\;.\\
c_4 \!\!&\!=\!&\!\! \frac{6c}{5}\frac{(h_{j+1}-h_j)}{h_{j+1}h_j}+\frac{m^2c^3}{2}(h_{j+1}+h_j)+\frac{1}{x_j}(\frac{Z^2}{c}-
c\kappa^2)(r_{j+1}h_{j+1}-r_{j-1}h_j)\;.\\
c_5 \!\!&\!=\!&\!\! -\frac{mcZ}{2x_j}(h_{j+1}+h_j)-mcZ(r_{j+1}h_{j+1}-r_{j-1}h_j)+\frac{6}{5}\frac{c\kappa}{x_j}
\frac{1}{h_{j+1}} (h_{j+1}-h_j)+c\kappa(r_{j-1}+\\
\!\!&\!\!&\!\! +r_j+r_{j+1})\;.\\
c_6 \!\!&\!=\!&\!\!  mc^2\frac{9}{70}(h_{j+1}-h_j)\;.\\
c_7 \!\!&\!=\!&\!\! -Z(2s_{j-1}+s_j)+mc^2\kappa s_{j-1}(h_{j+1}-h_j)+\frac{Z}{2x_j}(h_{j+1}+h_j)-
\frac{9}{70}\frac{mc^2\kappa}{x_j}h_{j+1}(h_{j+1}-h_j)\;.\\
c_8 \!\!&\!=\!&\!\! -\frac{9}{70c}h_{j+1}(h_{j+1}-h_j)\;.
\end{eqnarray*}
\begin{eqnarray*}
c_9 \!\!&\!=\!&\!\! -\frac{6}{5}\frac{1}{h_{j+1}}(h_{j+1}-h_j)\tau_j\;.\\
c_{10} \!\!&\!=\!&\!\!  \frac{\kappa}{2x_j}(h_{j+1}+h_j)\tau_j -\frac{Zs_{j-1}}{c}(h_{j+1}-h_j)+\kappa(r_{j+1}h_{j+1}-r_{j-1}h_j)\tau_j- \frac{9}{70}\frac{Z}{cx_j}h_{j+1}(h_{j+1}+\\
\!\!&\!\!&\!\! -h_j)\;.\\
c_{11} \!\!&\!=\!&\!\! -\frac{6mc^2}{5}\frac{1}{h_{j+1}}(h_{j+1}-h_j)\;.\\
c_{12} \!\!&\!=\!&\!\! -\frac{6}{5}\frac{Z}{x_j}\frac{1}{h_{j+1}}(h_{j+1}-h_j)+mc^2\kappa(r_{j+1}h_{j+1}- r_{j-1}h_j)-Z(r_{j-1}+r_j+r_{j+1})+\\
\!\!&\!\!&\!\! +\frac{\kappa mc^2}{2x_j}(h_{j+1}+h_j)\;.\\
c_{13} \!\!&\!=\!&\!\! \frac{9}{70}m^2c^3h_{j+1}(h_{j+1}-h_j)+\frac{c\kappa^2s_{j-1}}{x_j}(h_{j+1}-h_j)\;.\\
c_{14} \!\!&\!=\!&\!\! \frac{9}{70}\frac{mcZ}{x_j}h_{j+1}(h_{j+1}-h_j)+c\kappa(2s_{j-1}+s_j)- \frac{c\kappa}{2x_j}(h_{j+1}+h_j)\;.\\
c_{15} \!\!&\!=\!&\!\! -\frac{Zs_{j-1}}{c}(mc^2+\frac{Z}{x_j})(h_{j+1}-h_j)\;.\\
c_{16} \!\!&\!=\!&\!\! -\frac{Zs_{j-1}}{c}(mc^2-\frac{Z}{x_j})(h_{j+1}-h_j)\;.
\end{eqnarray*}

The following lemma provides the behavior of the eigenvalues in the vicinity of $x$ at infinity.
\begin{Lem}
Define the operator
\begin{equation*}
\mathcal{T}=\left(\begin{array}{cc} mc^2&-cD_x \\ cD_x&-mc^2\end{array}\right)\,.
\end{equation*}
Then for the radial Coulomb-Dirac equation
\begin{equation*}
\left(\mathcal{T}+\left(
\begin{array}{cc}
V(x) & c\frac{\kappa}{x} \\
c\frac{\kappa}{x} & V(x)
\end{array}
\right)\right)
 \left(
\begin{array}{c}
f(x) \\
g(x)
\end{array}
\right)
=
 \lambda \left(
\begin{array}{c}
f(x) \\
g(x)
\end{array}
\right)\, ,
\end{equation*}
the only accumulation point of the eigenvalues $\lambda$ is $mc^2$.\\

\hspace{-4mm}\underline{Proof}. \emph{ See \cite{GRI}.}\hfill{$\blacksquare$}
\end{Lem}
We now give the proof of the main theorem.\\
\underline{\emph{Proof}}. Consider the weak formulations (\ref{61}) and (\ref{62}), rewrite both of them as the following matrix-system
\begin{eqnarray}\label{66}
(mc^2-\lambda)M\negmedspace M_{000}\zeta-cM\negmedspace M_{010}\xi+c\kappa M\negmedspace M_{001}\xi-ZM\negmedspace M_{001}\zeta+c\tau M\negmedspace M_{110}\zeta+\\\nonumber
+c\kappa\tau M\negmedspace M_{101}\zeta-(mc^2+\lambda)\tau M\negmedspace M_{100}\xi-Z\tau M\negmedspace M_{101}\xi=0
\end{eqnarray}
and
\begin{eqnarray}
(mc^2-\lambda)\tau M\negmedspace M_{100}\zeta-c\tau M\negmedspace M_{110}\xi+c\kappa\tau M\negmedspace M_{101}\xi-Z\tau M\negmedspace M_{101}\zeta+cM\negmedspace M_{010}\zeta+\\\nonumber
+c\kappa M\negmedspace M_{001}\zeta-(mc^2+\lambda)M\negmedspace M_{000}\xi-ZM\negmedspace M_{001}\xi=0\, .
\end{eqnarray}

\vspace{7cm}

\hspace{-0.45cm}Where $\zeta=(\zeta_1,\cdots,\zeta_{j-1},\zeta_{j},\zeta_{j+1},\cdots ,\zeta_n,\zeta_1',\cdots,\zeta_{j-1}',\zeta_{j}',\zeta_{j+1}',\cdots ,\zeta_n')$ and $\xi=(\xi_1,\cdots,\xi_{j-1},\xi_{j},\\\xi_{j+1},\cdots ,\xi_n,\xi_1',\cdots,\xi_{j-1}',\xi_{j}',\xi_{j+1}',\cdots ,\xi_n')$. To get $\tau$ locally, that is $\tau_j$, for each subelement of the mesh, we consider the above equations for arbitrary $j$ cell. Employing Remark 2 and Remark 3 together with Lemma 2 we end up with
\begin{align}\label{68}
\Big(mc^2-\lambda\Big)\Big(\frac{9}{70}h_{j+1}\zeta_{j-1}+\frac{13}{35}(h_{j+1}+h_j)\zeta_j + \frac{9}{70}h_{j+1}\zeta_{j+1}\Big)-c\Big(\negthickspace -\frac{1}{2}\xi_{j-1}+\frac{1}{2}\xi_{j+1}\Big)+\\\nonumber
+c\kappa\Big(s_{j-1}\xi_{j-1}+s_{j}
\xi_{j}+s_{j-1}\xi_{j+1}\Big)-Z\Big(s_{j-1}\zeta_{j-1} +s_{j}\zeta_{j}+s_{j-1}\zeta_{j+1}\Big)+\\\nonumber
+\tau_j c\Big(-\frac{6}{5}\frac{1}{h_{j+1}}\zeta_{j-1}+\frac{6}{5}\frac{(h_{j+1}+h_j)}{h_{j+1}h_j}
\zeta_j -\frac{6}{5}\frac{1}{h_{j+1}}\zeta_{j+1}\Big)+\tau_j c\kappa \Big(r_{j-1}\zeta_{j-1} +r_{j}\zeta_{j}+r_{j+1}\zeta_{j+1}\Big)+\\\nonumber
-\tau_j\Big(mc^2+\lambda \Big)\Big(\frac{1}{2}\xi_{j-1}-\frac{1}{2}\xi_{j+1}\Big)-\tau_j Z\Big(r_{j-1}\xi_{j-1}+r_{j}
\xi_{j}+r_{j+1}\xi_{j+1}\Big)=0
\end{align}
and
\begin{align}\label{69}
\tau_j\Big(mc^2-\lambda \Big)\Big(\frac{1}{2}\zeta_{j-1}-\frac{1}{2}\zeta_{j+1}\Big)-\tau_j c\Big(-\frac{6}{5}\frac{1}{h_{j+1}}\xi_{j-1}+\frac{6}{5}\frac{(h_{j+1}+h_j)}{h_{j+1}h_j}\xi_j-\frac{6}{5}\frac{1}{h_{j+1}}\xi_{j+1}\Big)+\\\nonumber
 +\tau_j c\kappa \Big(r_{j-1}\xi_{j-1} +r_{j}\xi_{j}+r_{j+1}\xi_{j+1}\Big)-\tau_j Z\Big(r_{j-1}\zeta_{j-1}+r_{j}
\zeta_{j}+r_{j+1}\zeta_{j+1}\Big)+c\Big(\negthickspace-\frac{1}{2}\zeta_{j-1}+\frac{1}{2}\zeta_{j+1}\Big)+\\\nonumber
+c\kappa\Big(s_{j-1}\zeta_{j-1}+s_{j}
\zeta_{j}+s_{j-1}\zeta_{j+1}\Big)-\Big(mc^2+\lambda\Big)\Big(\frac{9}{70}h_{j+1}\xi_{j-1}+\frac{13}{35}(h_{j+1}+h_j)\xi_j +\\\nonumber
+ \frac{9}{70}h_{j+1}\xi_{j+1}\Big)-Z\Big(s_{j-1}\xi_{j-1} +s_{j}\xi_{j}+s_{j-1}\xi_{j+1}\Big)=0\, .
\end{align}
Using Lemma 1 (to substitute the nodal values $\zeta_{j-1}$, $\zeta_{j+1}$, $\xi_{j-1}$, and $\xi_{j+1}$), the equations above are written as
\begin{align}\label{70}
\Big(\frac{c}{2}+c\kappa s_{j-1} - \frac{1}{2}(mc^2+\lambda)\tau_j -Zr_{j-1}\tau_j \Big)\Big(\big(1-\frac{h_j\kappa}{x_j}\big)\xi_j+\big(\negthickspace-\frac{h_j}{c}(mc^2-
\frac{Z}{x_j})+\frac{h_j}{c}\lambda \big)\zeta_j \Big)+\\\nonumber
+\Big(\frac{13}{35}(h_{j+1}+h_j)(mc^2-\lambda)-Zs_{j}+\frac{6c}{5}\frac{(h_{j+1}+h_j)}{h_{j+1}h_j}
\tau_j+c\kappa r_j \tau_j\Big)\Big(\zeta_j\Big)+\Big(c\kappa s_j-Zr_j\tau_j\Big)\Big(\xi_j\Big)+  \\\nonumber
+\Big(\negthickspace-\frac{c}{2}+c\kappa s_{j-1} + \frac{1}{2}(mc^2+\lambda)\tau_j -Zr_{j+1}\tau_j \Big)\Big(\big(1+\frac{h_{j+1}\kappa}{x_j}\big)\xi_j+\big(\frac{h_{j+1}}{c}(mc^2-
\frac{Z}{x_j})-\frac{h_{j+1}}{c}\lambda \big)\zeta_j \Big)+\\\nonumber
+\Big(\frac{9}{70}h_{j+1}(mc^2-\lambda)-Zs_{j-1}-\frac{6c}{5}\frac{1}{h_{j+1}}\tau_j +c\kappa r_{j-1}\tau_j\Big)\Big( \big(1+\frac{h_j\kappa}{x_j}\big)\zeta_j+\big(\frac{h_j}{c}(-mc^2-\frac{Z}{x_j})+\\\nonumber
-\frac{h_j}{c}\lambda \big)\xi_j \Big)+\Big(\frac{9}{70}h_{j+1}(mc^2-\lambda)-Zs_{j-1}-\frac{6c}{5}\frac{1}{h_{j+1}}\tau_j+c\kappa r_{j+1}\tau_j\Big)\Big(\big(1-\frac{h_{j+1}\kappa}{x_j}\big)\zeta_j+\\\nonumber
+\big(\negthickspace-\frac{h_{j+1}}{c}(-mc^2-\frac{Z}{x_j})+\frac{h_{j+1}}{c}\lambda \big)\xi_j \Big)=0
\end{align}
and
\begin{align}\label{71}
 \Big(\negthickspace-\frac{c}{2}+c\kappa s_{j-1} + \frac{1}{2}(mc^2\!-\lambda)\tau_j\! -\!Zr_{j-1}\tau_j \Big)\Big( \big(1+\frac{h_j\kappa}{x_j}\big)\zeta_j+\big(\frac{h_j}{c}(-mc^2-\frac{Z}{x_j})-\frac{h_j}{c}\lambda \big)\xi_j \Big) +\\\nonumber
 +\Big(\negthickspace-\frac{13}{35}(h_{j+1}+h_j)(mc^2+\lambda)-Zs_{j}-\frac{6c}{5}
 \frac{(h_{j+1}+h_j)}{h_{j+1}h_j}\tau_j+c\kappa r_j \tau_j\Big)\Big(\xi_j\Big)+\Big(c\kappa s_j-Zr_j\tau_j\Big)\Big(\zeta_j\Big)+\\\nonumber
+ \Big(\frac{c}{2}+c\kappa s_{j-1} - \frac{1}{2}(mc^2-\lambda)\tau_j -Zr_{j+1}\tau_j \Big)\Big( \big(1-\frac{h_{j+1}\kappa}{x_j}\big)\zeta_j+\big(\negthickspace-\frac{h_{j+1}}{c}(-mc^2-
\frac{Z}{x_j}) +\\\nonumber
+\frac{h_{j+1}}{c}\lambda \big)\xi_j\Big)+ \Big(\negthickspace-\frac{9}{70}h_{j+1}(mc^2+\lambda)-Zs_{j-1}+\frac{6c}{5}\frac{1}{h_{j+1}}
\tau_j +c\kappa r_{j-1}\tau_j\Big)\Big( \big(1-\frac{h_j\kappa}{x_j}\big)\xi_j +\\\nonumber
+\big(\negthickspace-
\frac{h_j}{c}
(mc^2-\frac{Z}{x_j})+\frac{h_j}{c}\lambda \big)\zeta_j\Big) +\Big(\negthickspace-\frac{9}{70}h_{j+1}(mc^2+\lambda)-Zs_{j-1}+\frac{6c}{5}
 \frac{1}{h_{j+1}}\tau_j+\\\nonumber
+c\kappa r_{j+1}\tau_j\Big)\Big( \big(1+\frac{h_{j+1}\kappa}{x_j}\big)\xi_j +\big(\frac{h_{j+1}}{c}(mc^2-\frac{Z}{x_j})-\frac{h_{j+1}}{c}\lambda \big)\zeta_j \Big)=0\, .
\end{align}
Rewriting $(\ref{70})$ and $(\ref{71})$ by collecting the terms of $\zeta_j$ and of $\xi_j$ respectively gives
\begin{align}\label{72}
\Big[\Big(\frac{9}{70}h_{j+1}(mc^2-\lambda)-Zs_{j-1}-\frac{6c}{5}\frac{1}{h_{j+1}}\tau_j\Big)\Big( 2+\frac{\kappa}{x_j}(h_j-h_{j+1})\Big)+c\kappa\big(r_{j-1}+r_j+r_{j+1}+\\\nonumber
 +\frac{\kappa r_{j-1}}{x_j}h_j-\frac{\kappa r_{j+1}}{x_j}h_{j+1}  \big)\tau_j +\frac{6c}{5}\frac{(h_{j+1}+h_j)}{h_{j+1}h_j}\tau_j
+\frac{13}{35}(h_{j+1}+h_j)(mc^2-\lambda)-Zs_{j}+\Big(mc^2 +\\\nonumber
-\frac{Z}{x_j}-\lambda \Big)\Big(\negthickspace-\frac{h_j}{2}-\kappa s_{j-1}h_j+\frac{h_j}{2c}(mc^2+\lambda)\tau_j+\frac{Zr_{j-1}}{c}h_j\tau_j -\frac{h_{j+1}}{2}+\kappa s_{j-1}h_{j+1}+\frac{h_{j+1}}{2c}(mc^2 +\\\nonumber
+\lambda)\tau_j-\frac{Zr_{j+1}}{c}h_{j+1}\tau_j\Big)\Big]\zeta_j+\Big[\Big( mc^2+\frac{Z}{x_j}+\lambda\Big)\Big(\negthickspace-\frac{9}{70c}h_{j+1}h_j(mc^2-\lambda)+
\frac{Zs_{j-1}}{c}h_j +\\\nonumber
+\frac{6}{5}\frac{h_j}{h_{j+1}}\tau_j-\kappa r_{j-1}h_j\tau_j+\frac{9}{70c}h_{j+1}^2(mc^2-\lambda)-\frac{Zs_{j-1}}{c}h_{j+1}-
\frac{6}{5}\tau_j+\kappa r_{j+1}h_{j+1}\tau_j \Big)+\\\nonumber
+c\kappa(2s_{j-1}+s_j)-Z(r_{j-1}+r_j+r_{j+1})\tau_j-\frac{c\kappa}{2x_j}(h_j+h_{j+1})+
\frac{\kappa^2cs_{j-1}}{x_j} (h_{j+1}-h_j)+\\\nonumber
+\frac{\kappa}{2x_j}(mc^2+\lambda)(h_{j+1}+h_j)\tau_j+\frac{Zr_{j-1}\kappa}{x_j}h_j\tau_j-\frac{Zr_{j+1}
\kappa}{x_j}h_{j+1}\tau_j \Big]\xi_j=0
\end{align}
and

\begin{align}\label{73}
\Big[\!\!-\!Z(r_{j-1}\!+\!r_j\!+\!r_{j+1})\tau_j +c\kappa(2s_{j-1}+s_j)+\frac{\kappa}{2x_j}(mc^2\!-\!\lambda)(h_{j+1}+h_j)\tau_j
-\frac{c\kappa}{2x_j}(h_{j+1}+h_j)+\\\nonumber
+\frac{c\kappa^2s_{j-1}}{x_j}(h_j-h_{j+1})-\frac{Zr_{j-1}\kappa}{x_j}h_{j}\tau_j+
\frac{Zr_{j+1}\kappa}{x_j}h_{j+1}\tau_j+\Big( mc^2-\frac{Z}{x_j}-\lambda\Big)\Big(\frac{9}{70c}h_{j+1}h_j(mc^2+\lambda)+\\\nonumber
+\frac{Zs_{j-1}}{c}h_j - \frac{6}{5}\frac{h_j}{h_{j+1}}\tau_j
-\kappa r_{j-1}h_j\tau_j-\frac{9}{70c}h_{j+1}^2(mc^2+\lambda)-\frac{Zs_{j-1}}{c}h_{j+1}+
\frac{6}{5}\tau_j+\kappa r_{j+1}h_{j+1}\tau_j \Big) \Big]\zeta_j+\\\nonumber
+\Big[\Big(\negthickspace-\frac{9}{70}h_{j+1}(mc^2+\lambda)-Zs_{j-1}+\frac{6c}{5}\frac{1}{h_{j+1}}
\tau_j\Big)\Big( 2+\frac{\kappa}{x_j}(h_{j+1}-h_j)\Big)+c\kappa\big(r_{j-1}+r_j+r_{j+1}+\\\nonumber
-\frac{\kappa r_{j-1}}{x_j}h_j +\frac{\kappa r_{j+1}}{x_j}h_{j+1}  \big)\tau_j -\frac{6c}{5}\frac{(h_{j+1}+h_j)}{h_{j+1}h_j}\tau_j
-\frac{13}{35}(h_{j+1}+h_j)(mc^2+\lambda)-Zs_{j} +\cdots\nonumber
\end{align}
\begin{align*}
\cdots+\Big(mc^2+\frac{Z}{x_j}+\lambda \Big)\Big(\frac{h_j}{2}-\kappa s_{j-1}h_j-\frac{h_j}{2c}(mc^2-\lambda)\tau_j+\frac{Zr_{j-1}}{c}h_j\tau_j +\frac{h_{j+1}}{2} +\\\nonumber
+\kappa s_{j-1}h_{j+1}-\frac{h_{j+1}}{2c}(mc^2-\lambda)\tau_j-\frac{Zr_{j+1}}{c}h_{j+1}\tau_j\Big)\Big]\xi_j=0\, .\nonumber
\end{align*}

Gathering the factors of $\lambda^2$, $\lambda$, $\tau_j$, and the free terms in each equation for $\zeta_j$ and $\xi_j$ respectively, and using the defined above notations $c_i$'s, one can simplify $(\ref{72})$ and $(\ref{73})$ as follow
\begin{eqnarray}\label{741}
\Big[c_1\tau_j\lambda^2+(c_2+c_3)\lambda+(c_4+c_5)\tau_j+(c_6+c_7) \Big]\zeta_j+\Big[ c_8\lambda^2+(c_9+c_{10})\lambda+\\\nonumber
+(c_{11}+c_{12})\tau_j+(c_{13}+c_{14}+c_{15}) \Big]\xi_j=0
\end{eqnarray}
and
\begin{eqnarray}\label{742}
\Big[-c_8\lambda^2+(c_9-c_{10})\lambda+(-c_{11}+c_{12})\tau_j+(-c_{13}+c_{14}+c_{16})\Big]\zeta_j+\Big[ -c_1\tau_j\lambda^2+\\\nonumber
+(c_2-c_3)\lambda+(-c_4+c_5)\tau_j+(-c_6+c_7) \Big]\xi_j=0\, .
\end{eqnarray}

We consider the case where major part of the difficulty of solving the radial Dirac operators comes in. The above formulation is reduced to the operator $\mathcal{T}$ given in Lemma 3, the limit equation at infinity. One can understand the issue as the derived $\tau_j$ should guarantee the stability of the computations in the entire domain, particularly for large $x$, which is the operator $\mathcal{T}$ in one hand, and to consider the dominant part of the operator which causes the instability in the computations in the other. These motivations allow to consider $(\ref{741})$ and $(\ref{742})$ in the limit case
\begin{eqnarray}\label{75}
\Big[-\frac{(h_{j+1}+h_j)}{2c}\tau_j\lambda^2-\frac{9}{70}(h_{j+1}-h_j)\lambda+
\Big(\frac{6c}{5}\frac{(h_{j+1}-h_j)}{h_{j+1}h_j}+\frac{m^2c^3}{2}(h_{j+1}+h_j)\Big)\tau_j+\\\nonumber
+\frac{9}{70}mc^2(h_{j+1}-h_j) \Big]\zeta_j+\Big[\negthickspace-\negthickspace \frac{9}{70c}h_{j+1}(h_{j+1}-h_j)\lambda^2-\frac{6}{5}\frac{1}{h_{j+1}}(h_{j+1}-h_j)\tau_j\lambda
+\\\nonumber
-\frac{6mc^2}{5}\frac{1}{h_{j+1}}(h_{j+1}-h_j)\tau_j+\frac{9}{70}m^2c^3h_{j+1}(h_{j+1}-h_j) \Big]\xi_j=0
\end{eqnarray}
and
\begin{eqnarray}\label{76}
\Big[\frac{9}{70c}h_{j+1}(h_{j+1}-h_j)\lambda^2-\frac{6}{5}\frac{1}{h_{j+1}}(h_{j+1}-h_j)
\tau_j\lambda+\frac{6mc^2}{5}\frac{1}{h_{j+1}}(h_{j+1}-h_j)\tau_j+\\\nonumber
-\frac{9}{70}m^2c^3h_{j+1}(h_{j+1}-h_j)
\Big]\zeta_j+\Big[\frac{(h_{j+1}+h_j)}{2c}\tau_j\lambda^2
-\frac{9}{70}(h_{j+1}-h_j)\lambda+\Big(-\frac{6c}{5}\frac{(h_{j+1}-h_j)}{h_{j+1}h_j}+\\\nonumber
-\frac{m^2c^3}{2}(h_{j+1}+h_j)\Big)\tau_j - \frac{9}{70}mc^2(h_{j+1}-h_j)\Big]\xi_j=0\, .
\end{eqnarray}
Let $m=1$,  and define $\nabla_j=\frac{(h_{j+1}+h_j)}{(h_{j+1}-h_j)}$ and $\rho=-9/70$. Divide (\ref{75}) and (\ref{76}) by the quantity $h_{j+1}-h_j$ ($\neq 0$ for non-uniform mesh). In the vicinity of $c$ at infinity one gets the following dominant equations
\begin{equation}\label{77}
[\rho\lambda-a_j]\zeta_j+[d_j\lambda-b_j]\xi_j=0
\end{equation}
and
\begin{equation}\label{77half}
[d_j\lambda+b_j]\zeta_j+[\rho\lambda+a_j]\xi_j=0\, ,
\end{equation}
where
$
a_j=-\big(\frac{6c}{5}\frac{1}{h_{j+1}h_j}+\frac{c^3}{2}\nabla_j \big)\tau_j +\rho c^2
$,
$
b_j=\frac{6c^2}{5}\frac{1}{h_{j+1}}\tau_j-\frac{9c^3}{70}h_{j+1}
$,
and
$
d_j=-\frac{6}{5}\frac{1}{h_{j+1}}\tau_j\, .
$\\

Equations (\ref{77}) and (\ref{77half}) can be written as
\begin{equation}\label{79}
\left(
\begin{array}{cc}
\rho\lambda-a_j & d_j\lambda-b_j \\
d_j\lambda+b_j & \rho\lambda+a_j
\end{array}
\right)
 \left(
\begin{array}{c}
\zeta_j \\
\xi_j
\end{array}
\right)
=
  \left(
\begin{array}{c}
0 \\
0
\end{array}
\right)\, .
\end{equation}
Since $\zeta_j$ and $\xi_j$ are not identically zero for all $j$, then
\begin{equation}\label{80}
\left|
\begin{array}{cc}
\rho\lambda-a_j & d_j\lambda-b_j \\
d_j\lambda+b_j & \rho\lambda+a_j
\end{array}
\right|
 =0\, ,
\end{equation}
which gives
\begin{equation}\label{81}
\lambda_{1,2}=\pm\sqrt{(a_j^2-b_j^2)/(\rho^2-d_j^2)}\, .
\end{equation}
Since $c^2$ is the accumulation eigenvalue (Lemma 3, with $m=1$) we will only consider the positive part of $\lambda$ above named as $\lambda_1$. Now we would like to have $|\lambda_1-c^2|=0$
\begin{eqnarray*}
\begin{array}{lll}
                      \vspace{2mm}
                     & |\lambda_1-c^2|=0&\\
                     \vspace{2mm}
 \Longleftrightarrow & \frac{a_j^2-b_j^2}{c^4}=\rho^2-d_j^2&\\
                     \vspace{2mm}
 \Longleftrightarrow & c^4(\rho^2- \frac{36}{25}\frac{1}{h_{j+1}^2} \tau_j^2)& =\frac{36c^2}{25}\frac{1}{h_{j+1}^2h_j^2}\tau_j^2+\frac{c^6}{4}\nabla_j^2\tau_j^2+ \frac{6c^4}{5}\frac{1}{h_{j+1}h_j}\nabla_j\tau_j^2+\\
                      \vspace{2mm}
                     & &-\frac{12c^3}{5}\frac{1}{h_{j+1}h_j}\rho\tau_j-\rho c^5\nabla_j \tau_j-\frac{36c^4}{25}\frac{1}{h_{j+1}^2}\tau_j^2+\\
                     \vspace{2mm}
                     & &+\rho^2c^4+ \frac{54c^5}{175}\tau_j-\frac{81c^6}{4900}h_{j+1}^2\, ,
\end{array}
\end{eqnarray*}
keeping in mind the $c$ limit at infinity, the above formulation gives
\begin{equation}\label{82}
\frac{1}{4}\nabla_j^2\tau_j^2-\frac{81}{4900}h_{j+1}^2=0\, .
\end{equation}
The desired result is then obtained straight forward after substituting the value of $\nabla_j$ as defined before, and this ends the proof.\hfill{$\blacksquare$}\\

The derived $\tau$ provides complete cleaning of spectrum pollution for both categories. Also it is notable that the expression of $\tau$ treats the difficulty of the wave transferring between any two adjacent unbalanced mesh steps. The size of $\tau$ is proportional to the mesh size, i.e since we are dealing with exponentially distributed nodal points, $\tau$ has small size near the singularity $x=0$ due to the small mesh size, where it takes relatively large values in the region away from the origin which is dominated by coarse mesh.\\

Tables 7, 8, and 9 show the first computed energies for the electron in the Hydrogen-like Magnesium ion for both point and extended nucleus with $\kappa=|2|$. Table 7 shows the computed eigenvalues using the usual Galerkin formulation with linear basis functions. The number of interior nodal points used is $400$. Table 8 shows the same computations using the stability scheme. Table 9 represents the computed energies for extended nucleus using uniformly distributed charge with interior nodal points $397$, where $16$ nodal points are considered in the domain $[0\,,\,R]$ ($R$ is the radius of the nucleus).\\

\begin{table}[h]
\begin{normalsize}
\caption{The first computed eigenvalues for the electron in the Hydrogen-like Magnesium ion using usual FEM with linear basis functions for point nucleus.}
\centering
\begin{tabular}{@{} l c c r @{} }
\hline\hline
Level & $\kappa=2$ & $\kappa=-2$ & Rel. Form. $\kappa=-2$ \\ [0.5ex]
\hline
1  &\cellcolor[gray]{0.6} -18.0086349982 & -18.0086349982 & -18.0086349982 \\
2  & -8.00511829944 & -8.00511829944 & -8.00511739963 \\
3  & -4.50270135222 & -4.50270135225 & -4.50269856638 \\
$\Rrightarrow$ &\cellcolor[gray]{0.6} -2.88546212211 &\cellcolor[gray]{0.6} -2.88546212205 & Spurious Eigenvalue \\
4  & -2.88155295096 & -2.88155295095 & -2.88154739168 \\
5  & -2.00096852250 & -2.00096852249 & -2.00095939879 \\
6  & -1.47003410346 & -1.47003410350 & -1.47002066823 \\
$\Rrightarrow$ &\cellcolor[gray]{0.6} -1.13034880166 &\cellcolor[gray]{0.6} -1.13034880167 & Spurious Eigenvalue \\
7  & -1.12545691681 & -1.12545691683 & -1.12543844140 \\
8  & -.889228944495 & -.889228944484 & -.889204706429 \\
9  & -.720265553198 & -.720265553187 & -.720234829539 \\
$\Rrightarrow$ &\cellcolor[gray]{0.6} -.600492562625 &\cellcolor[gray]{0.6} -.600492562622 & Spurious Eigenvalue \\
10 & -.595258516248 & -.595258516277 & -.595220579682 \\
11 & -.500185771976 & -.500185772005 & -.500139887884 \\
12 & -.426201311278 & -.426201311300 & -.426146735771  \\ [1ex]
\hline
\end{tabular}
\label{table:nonlin}
\end{normalsize}
\end{table}

\begin{table}[h]
\begin{normalsize}
\caption{The first computed eigenvalues for the electron in the Hydrogen-like Magnesium ion using the stability scheme for point nucleus.}
\centering
\begin{tabular}{@{} l c c r @{} }
\hline\hline
Level & $\kappa=2$ & $\kappa=-2$ & Rel. Form. $\kappa=-2$ \\ [0.5ex]
\hline
1  &                   & -18.0086349985 & -18.0086349982 \\
2  & -8.00511739978    & -8.00511740020 & -8.00511739963 \\
3  & -4.50269856669    & -4.50269856719 & -4.50269856638 \\
4  & -2.88154739219    & -2.88154739270 & -2.88154739168 \\
5  & -2.00095939948    & -2.00095939991 & -2.00095939879 \\
6  & -1.47002066888    & -1.47002066924 & -1.47002066823 \\
7  & -1.12543844176    & -1.12543844201 & -1.12543844140 \\
8  & -.889204706068    & -.889204706109 & -.889204706429 \\
9  & -.720234827833    & -.720234827687 & -.720234829539 \\
10 & -.595220575840    & -.595220575531 & -.595220579682 \\
11 & -.500139880950    & -.500139880357 & -.500139887884 \\
12 & -.426146724530    & -.426146723650 & -.426146735771 \\
13 & -.367436809137    & -.367436807839 & -.367436826403 \\
14 & -.320073519367    & -.320073498169 & -.320073665658 \\
15 & -.281295132797    & -.281293164731 & -.281311119433 \\ [1ex]
\hline
\end{tabular}
\label{table:nonlin}
\end{normalsize}
\end{table}

\begin{table}[h]
\begin{normalsize}
\caption{The first computed eigenvalues for the electron in the Hydrogen-like Magnesium ion using the stability scheme for extended nucleus.}
\centering
\begin{tabular}{@{} l c c r @{} }
\hline\hline
Level & $\kappa=2$ & $\kappa=-2$ & Rel. Form. $\kappa=-2$ \\ [0.5ex]
\hline
1  &                & -18.0086349986 & -18.0086349982 \\
2  & -8.00511739975 & -8.00511740015 & -8.00511739963 \\
3  & -4.50269856673 & -4.50269856733 & -4.50269856638 \\
4  & -2.88154739230 & -2.88154739279 & -2.88154739168 \\
5  & -2.00095939956 & -2.00095940014 & -2.00095939879 \\
6  & -1.47002066903 & -1.47002066934 & -1.47002066823 \\
7  & -1.12543844179 & -1.12543844207 & -1.12543844140 \\
8  & -.889204706021 & -.889204706003 & -.889204706429 \\
9  & -.720234827640 & -.720234827433 & -.720234829539 \\
10 & -.595220575309 & -.595220574883 & -.595220579682 \\
11 & -.500139879906 & -.500139879215 & -.500139887884 \\
12 & -.426146722827 & -.426146721812 & -.426146735771 \\
13 & -.367436806543 & -.367436805088 & -.367436826403 \\
14 & -.320073514034 & -.320073492344 & -.320073665658 \\
15 & -.281294966822 & -.281292979627 & -.281311119433 \\ [1ex]
\hline
\end{tabular}
\label{table:nonlin}
\end{normalsize}
\end{table}

To study the convergence property of the derived scheme, we compare the approximated eigenvalues of the electron in the Hydrogen-like Magnesium ion for point nucleus using the usual FEM as in Table 7, to those values obtained by the stability scheme as in Table 8. Ignoring the presence of the spurious values, one notes that the relative error in the approximation of the first $12$ genuine eigenvalues using FEM is nearly $10^{-4}$. Whereas the relative error for the same group of eigenvalues using the stability scheme is not exceeding $3*10^{-8}$. Thus, the speed of convergence is also enhanced.\\

In Table 10, we provide the approximated eigenvalues for the electron in the Hydrogen-like Uranium ion using the stability scheme. The computations are obtained for different values of the quantum number $\kappa$ for extended nucleus. The number of nodal points used is 203 (13 out of them are used to discretize the segment $[0\,,\,R]$).\\

\textbf{Conclusion.}

Our computations indicate that the SUPG scheme applied to solve the radial Dirac eigenvalue problem is stable in the sense of complete elimination of spectrum pollution. This approach is mainly compiled of two strategies; the first is the suitable choice of the trial functional space. The second is based on varying the test function to live in another space different from that for the trial function, this strongly depends on the derived stability parameter $\tau$. The derived $\tau$ is a considerable achievement where its formula is rather easy to implement, and it yields full treatment of the spuriousity for both categories.
\newpage
\begin{table}[h]
\begin{footnotesize}
\caption{The first computed eigenvalues for the electron in the Hydrogen-like Uranium ion for different energy levels using the stability scheme for extended nucleus.}
\centering
\begin{tabular}{@{} l c c c c r @{} }
\hline\hline
Level & $\kappa=-1$ & $\kappa=1$ & $\kappa=-2$ &  $\kappa=2$ & $\kappa=-3$ \\ [0.5ex]
\hline
1& -4853.62949434& & & &\\
2& -1255.95827216&-1257.22738641& & &\\
3& -538.661380908& -539.033990526& -1089.61141552& &\\
4& -295.078728020& -295.232044507& -489.037085134& -489.037084960&\\
5& -185.395090636& -185.471947843& -274.407758128& -274.407757668& -476.261594535\\
6& -127.042256989& -127.086006093& -174.944613694& -174.944613207& -268.965877806\\
7& -92.4088112704& -92.4360075180& -121.057538281& -121.057537866& -172.155252828\\
8& -70.2043012114& -70.2223336849& -88.6717487653& -88.6717484812& -119.445272665\\
9& -55.1286483910& -55.1412076654& -67.7178951387& -67.7178950309& -87.6582879582\\
10& -44.4301782764&-44.4392710290& -53.3922002629& -53.3922003729& -67.0402332769\\
11& -36.5662117804& -36.5730039895& -43.1702540865& -43.1702544560& -52.9170997410\\
12& -30.6178633663& -30.6230696251& -35.6233695209& -35.6233701925& -42.8244637407\\
13& -26.0103096494& -26.0143875052& -29.8940993552& -29.8941003747& -35.3639479395\\
14& -22.3691011929& -22.3723545239& -25.4427187732& -25.4427201886& -29.6945373867\\
15& -19.4418733070& -19.4445102954& -21.9158181718& -21.9158200337& -25.2859399425\\
16& -17.0535375600& -17.0557046811& -19.0741660324& -19.0741683944& -21.7904231350\\
17& -15.0795424863& -15.0813452131& -16.7511595194& -16.7511624386& -18.9723111918\\
18& -13.4293342440& -13.4308500831& -14.8278955863& -14.8278991221& -16.6673054165\\
19& -12.0358025118& -12.0370894509& -13.2176791341& -13.2176833488& -14.7580403886\\
20& -10.8483590654& -10.8494611874& -11.8560968516& -11.8561018101& -13.1588737727\\
21& -9.82828760857& -9.82923891029& -10.6944840991& -10.6944898690& -11.8061295095\\
22& -8.94555154162& -8.94637858292& -9.69552155807& -9.69552820876& -10.6516697044\\
23& -8.17655984194& -8.17728361748& -8.83020376778& -8.83021137096& -9.65855817575\\
24& -7.50257602554& -7.50321330929& -8.07571205220& -8.07572068159& -8.79807266133\\
25& -6.90856705387& -6.90913137985& -7.41389618253& -7.41390591378& -8.04760687717\\
26& -6.38235847787& -6.38286086464& -6.83017355332& -6.83018446387& -7.38917199781\\
27& -5.91400613762& -5.91445563130& -6.31271965872& -6.31273182756& -6.80830955954\\
28& -5.49532323801& -5.49572732274& -5.85186500024& -5.85187850842& -6.29329159563\\
29& -5.11952039156& -5.11988530441& -5.43964039565& -5.43965532594& -5.83452444744\\
30& -4.78092881330& -4.78125978717& -5.06943037836& -5.06944681572& -5.42409906976\\
31& -4.47478541851& -4.47508687212& -4.73570629156& -4.73572432309& -5.05544809643\\
32& -4.19706449375& -4.19734018037& -4.43381880272& -4.43383851763& -4.72308165346\\
33& -3.94434475273& -3.94459787924& -4.15983518735& -4.15985667790& -4.42238191278\\
34& -3.71370352567& -3.71393684674& -3.91041067459& -3.91043403573& -4.14944192144\\
35& -3.50263193199& -3.50284782636& -3.68268594092& -3.68271127049& -3.90093812765\\
36& -3.30896641306& -3.30916694458& -3.47420485278& -3.47423225164& -3.67402878591\\
37& -3.13083310725& -3.13102007468& -3.28284801467& -3.28287758577& -3.46627240831\\
38& -2.96660233349& -2.96677731307& -3.10677874375& -3.10681058942& -3.27556186817\\
39& -2.81485102305& -2.81501541978& -2.94439883084& -2.94443304812& -3.10007081630\\
40& -2.67433187309& -2.67448701724& -2.79431185210& -2.79434854846& -2.93820980064\\
41& -2.54395104802& -2.54409831318& -2.65529221336& -2.65533164690& -2.78858986482\\
42& -2.42276031374& -2.42290074014& -2.52626102582& -2.52630405138& -2.64999168906\\
43& -2.30995095161& -2.31008292285& -2.40627974326& -2.40632804795& -2.52134098372\\
44& -2.20475351060& -2.20486906775& -2.29457262587& -2.29462526625& -2.40170120650\\
45& -2.10615086328& -2.10624270502& -2.19049573109& -2.19053698126& -2.29029930213\\ [1ex]
\hline
\end{tabular}
\label{table:nonlin}
\end{footnotesize}
\end{table}

\end{document}